\documentclass[prb,twocolumn,showpacs]{revtex4}

\usepackage{graphicx,epsf}
\usepackage{bm}

\begin{document}

\title{Enhancement of the Two-channel Kondo Effect in Single-Electron
       boxes}
\smallskip

\author{Eran Lebanon and Avraham Schiller}
\affiliation{Racah Institute of Physics, The Hebrew University,
             Jerusalem 91904, Israel}
\author{Frithjof B. Anders}
\affiliation{Institut f\"ur Festk\"orperphysik, Technical University
             Darmstadt, 64289 Darmstadt, Germany}

\begin{abstract}
The charging of a quantum box, coupled to a lead by tunneling
through a single resonant level, is studied near the degeneracy
points of the Coulomb blockade. Combining Wilson's numerical
renormalization-group method with perturbative scaling
approaches, the corresponding low-energy Hamiltonian is solved
for arbitrary temperatures, gate voltages, tunneling rates, and
energies of the impurity level. Similar to the case of a weak
tunnel barrier, the shape of the charge step is governed at
low temperatures by the non-Fermi-liquid fixed point of the
two-channel Kondo effect. However, the associated Kondo
temperature $T_K$ is strongly modified. Most notably, $T_K$
is proportional to the width of the level if the transmission
through the impurity is close to unity at the Fermi energy,
and is no longer exponentially small in one over the tunneling
matrix element. Focusing on a particle-hole symmetric level,
the two-channel Kondo effect is found to be robust against the
inclusion of an on-site repulsion on the level. For a large
on-site repulsion and a large asymmetry in the tunneling rates
to box and to the lead, there is a sequence of Kondo effects:
first the local magnetic moment that forms on the level undergoes
single-channel screening, followed by two-channel overscreening
of the charge fluctuations inside the box.
\end{abstract}



\maketitle

\section{Introduction}

The two-channel Kondo effect~\cite{NB80,Zawadowski80} is a
prototype for non-Fermi-liquid behavior in correlated electron
systems. It occurs when a spin-$\frac{1}{2}$ local moment is
coupled antiferromagnetically to two identical, independent
conduction-electron channels. Below a characteristic energy scale,
$k_B T_K$, the system is governed by an intermediate-coupling
non-Fermi-liquid fixed point, representing the fact that neither
conduction-electron channel can exactly screen the impurity
moment. The resulting low-energy physics is characterized by
anomalous thermodynamic and dynamic properties.~\cite{CZ98}
Hampering the quest for an experimental realization of the
two-channel Kondo effect is the extreme instability of the
non-Fermi-liquid fixed point against various perturbations.
Any channel asymmetry, however small, drives the system to
a Fermi-liquid fixed point, as does the application of a
magnetic field.~\cite{CZ98} Hence the observation of a fully
developed two-channel Kondo effect appears hopeless, unless one
is able to identify a system where the equivalence of the
two conduction-electron channels is guaranteed by symmetry,
and all relevant perturbations, such as an applied magnetic
field, can be tuned to zero.

One of the leading scenarios for the realization of the
two-channel Kondo effect is that of a quantum box, either a
small metallic grain or a large semiconducting quantum dot,
weakly connected to a lead by a single-mode point contact.
Near the degeneracy points of the Coulomb-blockade staircase,
one can map the charge fluctuations in the quantum box onto
a planner two-channel Kondo Hamiltonian,~\cite{Matveev91}
with the two available charge configurations in the box
playing the role of the impurity spin, and the physical
spin of the conduction electrons acting as a
passive channel index. The energy difference between the
two charge configurations corresponds in this mapping to
an effective magnetic field, which can be tuned to zero
by varying the gate voltage. Indeed, some signatures of the
two-channel Kondo effect were recently observed for such
a setting in semiconductor quantum dots.~\cite{BZAS99}

However, as recently emphasized by Zar\'and {\em et al.},~\cite{ZZW00}
measurement of the low-temperature, non-Fermi-liquid regime
of the two-channel Kondo effect sets opposite constraints
on the size of the quantum box. On the one hand, the charging
energy must be sufficiently large in order for a measurable
Kondo temperature to emerge, limiting the box from being
too large. On the other hand, the mean level spacing in the
box must be sufficiently small compared to $k_B T_K$, as not
to cut off the approach to the non-Fermi-liquid fixed point.
Hence the box cannot be too small. As argued by Zar\'and
{\em et al.},~\cite{ZZW00} these conflicting limitations
cannot be simultaneously realized in present-day semiconducting
devices. The alternate possibility of using metallic grains
is faced with a different difficulty of fabricating stable
atomic-size contacts, which are required for obtaining a
measurable $T_K$.~\cite{ZZW00} Hence the prospects for
obtaining a fully developed two-channel Kondo effect within
Matveev's original picture remain unclear.

In this paper we show that the two-channel Kondo temperature
$T_K$, and thus the chances for observing the two-channel
Kondo effect, can be greatly enhanced if tunneling between
the lead and the box takes place via a single resonant level.
The study of such resonant tunneling was initiated by
Gramespacher and Matveev,~\cite{GM00} who showed that one
can have a nearly perfect Coulomb staircase, even if the
transmission coefficient through the impurity is one at the
Fermi energy. This differs markedly from the case of an
energy-independent transmission coefficient, where
the Coulomb staircase is washed out for perfect
transmission.~\cite{Matveev95} Here we resolve the shape
of the Coulomb step separating two neighboring charge
plateaus, for the case of tunneling through a resonant
level.

Using combined analytical and numerical techniques we find
that the shape of the step is governed at low temperatures
by the non-Fermi-liquid fixed point of the two-channel Kondo
effect, similar to the case of a weak tunnel barrier.~\cite{Matveev91}
However, the associated Kondo temperature is strongly modified.
Most notably, $T_K$ is no longer exponentially small in
one over the tunneling matrix element if the transmission
through the impurity is close to unity at the Fermi energy,
but rather is proportional to the width of the level.
In general, $T_K$ strongly depends on the ratio
of the tunneling rates to the box and to the lead, which
illustrates the inequivalent roles of the two rates.
If the level is at resonance with the Fermi energy,
this ratio defines the crossover from weak to strong coupling.
The dependences of $T_K$ on the tunneling rates and
on the energy of the level are analyzed in detail,
as are the position and shape of the charge step.

A potential concern with the above setting has to do with
the effect of an on-site Coulomb repulsion on the impurity
level, as it couples the two spin channels. Modeling the
interacting level by a symmetric Anderson impurity, we
show that the two-channel Kondo effect is robust against
the inclusion of a Coulomb repulsion on the impurity level,
and that $T_K$ is enhanced by a moderately large repulsion
in the mixed-valent regime. For a large on-site repulsion,
a local magnetic moment is formed on the level. In this
regime, $T_K$ decays exponentially with on one over the
tunneling rates.

The remainder of the paper is organized as follows:
Section~\ref{sec:model} introduces the physical setting
under consideration. The relation with the two-channel Kondo
Hamiltonian is clarified in sec.~\ref{sec:two-channel-H},
for the case of a noninteracting level. An analytic
treatment of the weak-coupling regime for a noninteracting
level is presented in sec.~\ref{sec:weak_coupling}, both
for a level at resonance and a level off resonance with
the Fermi energy. This is followed in
sec.~\ref{sec:General_coupling} by a comprehensive
analysis of all coupling regimes using the numerical
renormalization-group method. The effect of an on-site
repulsion on the level is studied in turn in
sec.~\ref{sec:Interacting_level}, followed by
a discussion and a summary of our results in
sec.~\ref{sec:discussion}.

\begin{figure}
\centerline{
\includegraphics[width=75mm]{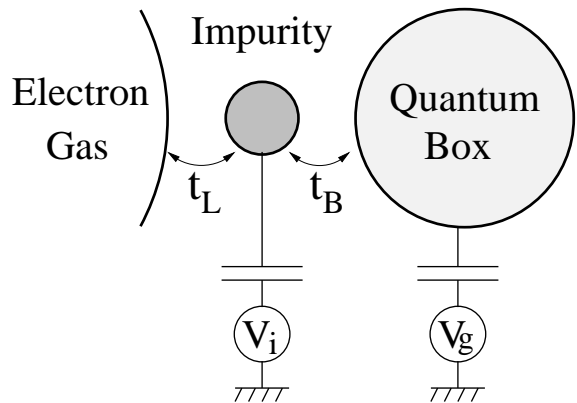}
}\vspace{0pt}
\caption{Schematic sketch of the physical setting. A quantum
         box and a metallic lead are each coupled by tunneling
	 to an intermediate impurity level. The charge inside
	 the quantum box is tuned by varying the gate voltage
	 $V_g$, while the energy of the level is controlled
	 by varying the gate voltage $V_i$.}
\label{fig:fig1}
\end{figure}

\section{Coulomb blockade with resonant tunneling}
\label{sec:model}

The physical setting under consideration is shown schematically
in Fig.~\ref{fig:fig1}. It consists of a metallic lead and a
quantum box, each coupled by tunneling to an impurity placed
in between the lead and box. The impurity is assumed to have
just a single energy level $\epsilon_d$ in the relevant energy
range, described by the two creation operators
$d^{\dagger}_{\uparrow}$ and $d^{\dagger}_{\downarrow}$. The
quantum box is characterized by the single-particle dispersion
$\epsilon_{k B}$, and by the charging energy $E_C = e^2/2C$.
Here $C$ is the capacitance of the box. The mean level spacing
inside the box is assumed to be considerably smaller than all
other energy scales in the problem, such that a continuum-limit
description can be used. The charge inside the box is controlled by
varying the gate voltage $V_g$, which determines the electrostatic
potential inside the box. We parameterize the latter by the
dimensionless number $N = C_g V_g/e$, where $C_g$ is the
capacitance of the gate, and $-e$ is the electron charge.

Modeling the lead by a noninteracting band with dispersion
$\epsilon_{k L}$, the Hamiltonian of the system is given by
\begin{eqnarray}
{\cal H} &=& \sum_{\alpha = L, B}\sum_{k,\sigma} \epsilon_{k \alpha}
             c^{\dagger}_{k \alpha \sigma} c_{k \alpha \sigma}
          + \epsilon_d \sum_{\sigma} d^{\dagger}_{\sigma} d_{\sigma}
\label{Full_H}
\\
&+& \sum_{k, \sigma \alpha} t_{k \alpha} \left \{
        c^{\dagger}_{k \alpha \sigma} d_{\sigma} + {\rm H.c.} \right \}
+ E_C ( \hat{n} - N )^2 ,
\nonumber
\end{eqnarray}
where $c^{\dagger}_{k L \sigma}$ ($c^{\dagger}_{k B \sigma}$)
creates an electron with spin projection $\sigma$ in the lead
(box); $t_{k L}$ ($t_{k B}$) are the matrix elements for
tunneling between the impurity and the lead (box); and
\begin{equation}
\hat{n} = \sum_{k, \sigma} \left [
                c^{\dagger}_{k B \sigma } c_{k B \sigma}
                - \theta(-\epsilon_{k B}) \right ]
\end{equation}
measures the number of excess electrons in the box. Here and
throughout the paper we set the chemical potential as our
reference energy, namely, all single-particle energies are
measured relative to the chemical potential. For an
interacting level, Eq.~(\ref{Full_H}) is supplemented by
the on-site repulsion term
\begin{equation}
{\cal H}_U = U \hat{n}_{d \uparrow} \hat{n}_{d \downarrow} ,
\label{Hubbard}
\end{equation}
where $\hat{n}_{d \sigma} = d^{\dagger}_{\sigma} d_{\sigma}$
are the number operators on the level.

The Hamiltonian of Eq.~(\ref{Full_H}) features two basic
energy scales,
\begin{eqnarray}
\Gamma_L &=& \pi \sum_{k} t_{k L}^2 \delta(\epsilon_{k L}),
\label{Gamma_L} \\
\Gamma_B &=& \pi \sum_{k} t_{k B}^2 \delta(\epsilon_{k B}),
\label{Gamma_B}
\end{eqnarray}
corresponding to half the tunneling rates from the impurity to
the lead and to the box, respectively. As shown by Gramespacher
and Matveev,~\cite{GM00} for $\Gamma_L, \Gamma_B \ll E_C$ there
is a nearly perfect Coulomb staircase, even if the transmission
coefficient through the impurity is one at the Fermi energy.
The shape of the sharp steps near half-integer values of $N$
was left unresolved in Ref.~\onlinecite{GM00}, which
is the objective of the present paper. To this end, we focus
hereafter on $N = n + \frac{1}{2} + \delta N$, where $n$ is an
integer and $|\delta N| \ll 1$. This range in $N$ corresponds
to the step separating the two charge plateaus with $n$ and
$n+1$ excess electrons in the box.

For temperatures well below the charging energy, $k_B T
\ll E_C$, only the $n$ and $n+1$ charge configurations are
thermally accessible in the box. Hence one can remove all
excited charge configurations by projecting the Hamiltonian of
Eq.~(\ref{Full_H}) onto the $n$ and $n+1$ subspaces. Following
Matveev,~\cite{Matveev91} a spin-$\frac{1}{2}$ isospin
operator $\vec{S}$ is introduced to label the two available
charge configurations: $S_z = 1/2$ for the $n+1$ subspace,
and $S_z = -1/2$ for the $n$ subspace. The raising and lowering
operators, $S^{\pm} = S_x \pm i S_y$, describe then transitions
between the $n$ and $n+1$ subspaces, corresponding to the
addition or removal of a box electron. Hence the Hamiltonian
of Eq.~(\ref{Full_H}) is converted to
${\cal H} = {\cal H}_L + {\cal H}_B + {\cal H}_{tun}$, where
\begin{eqnarray}
{\cal H}_L &=& \sum_{k,\sigma} \epsilon_{k L}
             c^{\dagger}_{k L \sigma} c_{k L \sigma}
          + \epsilon_d \sum_{\sigma} d^{\dagger}_{\sigma} d_{\sigma}
\label{H_L} \\
&+& \sum_{k, \sigma} t_{k L} \left \{
        c^{\dagger}_{k L \sigma} d_{\sigma} + {\rm H.c.} \right \}
\nonumber
\end{eqnarray}
describes the coupled lead and impurity,
\begin{equation}
{\cal H}_B = \sum_{k,\sigma} \epsilon_{k B}
             c^{\dagger}_{k B \sigma} c_{k B \sigma} - e V_B S_z
\label{H_B}
\end{equation}
with $e V_B = 2 E_C \delta N$ describes the isolated box, and
\begin{equation}
{\cal H}_{tun} = \sum_{k, \sigma} t_{k B} \left \{
        c^{\dagger}_{k B \sigma} d_{\sigma} S^{+}
        + {\rm H.c.} \right \}
\label{H_tun}
\end{equation}
describes tunneling between the impurity and the box.
For an interacting level, the above Hamiltonian is
supplemented by the on-site interaction term of
Eq.~(\ref{Hubbard}).

Equations~(\ref{H_L})--(\ref{H_tun}) are a straightforward
generalization of Matveev's original mapping for a weak
tunnel barrier.~\cite{Matveev91} As in the case of a weak
tunnel barrier, the average excess charge in the box takes
the form
\begin{equation}
\langle Q \rangle = -e \left( n + \frac{1}{2} \right)
        - e \langle S_z \rangle ,
\end{equation}
while the capacitance of the junction,
\begin{equation}
C(V_B, T) = -\partial \langle Q \rangle/\partial V_B,
\end{equation}
is proportional to the isospin susceptibility. The connection
between the Hamiltonian of Eqs.~(\ref{H_L})--(\ref{H_tun})
and the two-channel Kondo model is less transparent than
for a weak tunnel barrier, since the tunneling
Hamiltonian ${\cal H}_{tun}$ involves the localized
$d^{\dagger}_{\sigma}$ degrees of freedom. As we show
in the following section, one can still relate the
Hamiltonian of Eqs.~(\ref{H_L})--(\ref{H_tun}) to the
two-channel Kondo model, by first diagonalizing the
quadratic Hamiltonian term ${\cal H}_L$. This gives
rise to a new variant of the planner two-channel Kondo
Hamiltonian, in which the spin-up and the spin-down
conduction electrons have two distinct bandwidths.

\section{Relation to the two-channel Kondo Hamiltonian}
\label{sec:two-channel-H}

Much of the underlying physics of the Hamiltonian of
Eqs.~(\ref{H_L})--(\ref{H_tun}) can be understood by
converting to a single-particle basis that diagonalizes
the quadratic Hamiltonian term ${\cal H}_L$. The
objective of this section is to construct such
a basis. In doing so we assume that the level
$\epsilon_d$ lies well within the band (i.e., no
bound state is formed), and neglect for simplicity
all $k$-dependence of the tunneling matrix elements
$t_{k L}$ and $t_{k B}$. The latter are taken for
convenience to be real and positive.

A convenient representation of the eigen modes of
${\cal H}_L$ involves the $d$-electron Green function
\begin{equation}
G(z) = \left [ z - \epsilon_d - t_L^2 \sum_{k}
                \frac{1}{z - \epsilon_{k L} } \right]^{-1},
\label{eqn:d-electron}
\end{equation}
along with the associated phases
\begin{equation}
\phi_k = {\rm arg} \left \{ G(\epsilon_{k L} - i\eta) \right \} .
\end{equation}
Here $\eta$ is a positive infinitesimal. Introducing
the properly normalized fermion operators
\begin{eqnarray}
\psi^{\dagger}_{k L \sigma} =&& e^{i\phi_k} c^{\dagger}_{k L \sigma}
       + t_L |G(\epsilon_{k L} + i\eta)|
\\
&& \times \left [ d^{\dagger}_{\sigma} + t_L \sum_{k'}
       \frac{1}{\epsilon_{k L} - \epsilon_{k' L} + i\eta}
       c^{\dagger}_{k' L \sigma} \right ] ,
\nonumber
\end{eqnarray}
the Hamiltonian term ${\cal H}_L$ acquires the diagonal form
\begin{equation}
{\cal H}_L = \sum_{k,\sigma} \epsilon_{k L}
      \psi^{\dagger}_{k L \sigma} \psi_{k L \sigma} ,
\end{equation}
while the $d^{\dagger}_{\sigma}$ operators are expanded as
\begin{equation}
d^{\dagger}_{\sigma} = t_L \sum_{k} |G(\epsilon_{k L} + i\eta)|
      \psi^{\dagger}_{k L \sigma} .
\end{equation}
Further converting to the constant-energy-shell operators
\begin{eqnarray}
a^{\dagger}_{\epsilon L \sigma} &=&
      \frac{1}{\sqrt{\rho_{L}(\epsilon)}}
      \sum_k \delta(\epsilon - \epsilon_{k L} )
      \psi^{\dagger}_{k L \sigma} , \\
a^{\dagger}_{\epsilon B \sigma} &=&
      \frac{1}{\sqrt{\rho_{B}(\epsilon)}}
      \sum_k \delta(\epsilon - \epsilon_{k B} )
      c^{\dagger}_{k B \sigma}
\end{eqnarray}
[here $\rho_{L} (\epsilon)$ and $\rho_{B} (\epsilon)$ are
the underlying lead and box density of states, respectively],
the full Hamiltonian reads
\begin{eqnarray}
{\cal H} &=& \sum_{\alpha = L, B} \sum_{\sigma} \int
          \epsilon a^{\dagger}_{\epsilon \alpha \sigma}
          a_{\epsilon \alpha \sigma} d\epsilon - e V_B S_z
\label{H_energy}\\
&+& \frac{1}{2}\sum_{\sigma}
         \int d\epsilon \int d\epsilon' J(\epsilon, \epsilon')
         \left \{
                a^{\dagger}_{\epsilon B \sigma} a_{\epsilon' L \sigma}
                S^{+} + {\rm H.c.} \right \} .
\nonumber
\end{eqnarray}
Here, $e V_B$ is equal to $2 E_C \delta N$; the energy-dependent
coupling $J(\epsilon, \epsilon')$ is given by
\begin{equation}
J(\epsilon, \epsilon') = 2 t_L t_B
         \sqrt{ \rho_{B}(\epsilon) \rho_{L}(\epsilon') }
         |G(\epsilon' + i\eta)| ;
\label{J_ee'}
\end{equation}
and the single-particle operators $a_{\epsilon \alpha \sigma}$
obey canonical anticommutation relations:
\begin{equation}
\left \{ a_{\epsilon \alpha \sigma},
         a^{\dagger}_{\epsilon' \alpha' \sigma'} \right \} =
      \delta(\epsilon - \epsilon') \delta_{\alpha \alpha'}
         \delta_{\sigma \sigma'} .
\end{equation}
Note that we have omitted in Eq.~(\ref{H_energy}) all those
conduction-electron channels in both ${\cal H}_L$ and
${\cal H}_B$ that decouple from the tunneling term
${\cal H}_{tun}$.

Equation~(\ref{H_energy}) should be compared with the
corresponding constant-energy-shell representation of
the planner two-channel Kondo model with a local magnetic
field. In the latter case, the indices $L$ and $B$ are
replaced with spin-up and spin-down labels, $e V_B$
corresponds to the local magnetic field, and $\sigma$
acts as the channel index. More significantly,
$J(\epsilon, \epsilon')$ of Eq.~(\ref{J_ee'})
is replaced with $J_{2CK}(\epsilon, \epsilon') =
\sqrt{ \rho(\epsilon) \rho(\epsilon') } J_{\perp}$,
where $J_{\perp}$ is the transverse Kondo coupling, and
$\rho(\epsilon)$ is the joint density of states (DOS)
of the spin-up and spin-down conduction electrons. Thus,
identifying the lead and box indices $L$ and $B$ with
isospin-up and isospin-down labels,~\cite{Matveev91}
Eq.~(\ref{H_energy}) coincides with the planner two-channel
Kondo Hamiltonian modulo one crucial difference: the
effective DOS for the isospin-up and isospin-down
conduction electrons are markedly different in
Eq.~(\ref{H_energy}). While the isospin-down DOS is
equal to $\rho_B(\epsilon)$, the effective isospin-up
DOS is given by the spectral part of $G(z)$:~\cite{comment_on_rho_L^eff}
\begin{equation}
\rho_{L}^{\rm eff}(\epsilon) = -\frac{1}{\pi} {\rm Im}
         \left \{ G(\epsilon + i\eta) \right \} =
         t^2_L \rho_{L}(\epsilon) |G(\epsilon + i\eta)|^2 .
\label{rho_L_eff}
\end{equation}

In the wide-band limit, and for $\epsilon$ well within
the band, Eq.~(\ref{rho_L_eff}) reduces to the Lorentzian
form
\begin{equation}
\rho_{L}^{\rm eff}(\epsilon) = \frac{1}{\pi} \frac{\Gamma_L}
         { (\epsilon - \epsilon_d)^2 + \Gamma_L^2 } .
\label{Lorentzian}
\end{equation}
Hence the effective bandwidth for the isospin-up electrons
is equal to $\Gamma_L$, i.e., notably smaller than the
isospin-down bandwidth $D$.~\cite{comment_on_D} Such a large
separation of bandwidths for the isospin-up and isospin-down
conduction electrons has no analog in the conventional
two-channel Kondo Hamiltonian, where the two spin
orientations are identical for a zero magnetic field.
Moreover, $\rho_{L}^{\rm eff}(\epsilon)$ is centered about
$\epsilon_d$, which corresponds for $|\epsilon_d| \gg \Gamma_L$
either to a nearly filled band ($\epsilon_d < 0$)
or to a nearly empty band ($\epsilon_d > 0$).
Below we explore in detail the consequences of these
deviations from the conventional two-channel Kondo
Hamiltonian, but first let us give some heuristic arguments
for the expected low-energy physics in the case where
$\epsilon_d = 0$.

As is well known, the low-energy physics of the planner
two-channel Kondo model is governed by the dimensionless
coupling $\rho_0 J_{\perp}$, where $\rho_0 = \rho(0)$ is
the conduction-electron DOS at the Fermi energy. By analogy
with the two-channel Kondo Hamiltonian,
\begin{equation}
J(0,0) = \frac{2}{\pi} \sqrt{ \frac{\Gamma_L \Gamma_B}
                            { \epsilon_d^2 + \Gamma_L^2} }
\label{J_0_0}
\end{equation}
plays the role of $\rho_0 J_{\perp}$ in the Hamiltonian
of Eq.~(\ref{H_energy}). Specifically, for a level at
resonance with the Fermi energy, i.e., $\epsilon_d = 0$,
Eq.~(\ref{J_0_0}) is equal to $(2/\pi) \sqrt{\Gamma_B/\Gamma_L}$.

To the extent that the Hamiltonian of Eq.~(\ref{H_energy})
still flows for $\epsilon_d = 0$ and $V_B = 0$ to the
intermediate-coupling fixed point of the two-channel Kondo
effect, different qualitative behaviors are expected
of the Kondo temperature $T_K$ in each of the limits
$\Gamma_B \ll \Gamma_L$ and $\Gamma_L \ll \Gamma_B$.
For $\Gamma_B \ll \Gamma_L$, the bare spin-exchange
interaction is weak. Hence $T_K$ should be exponentially
small in $\sqrt{\Gamma_L/\Gamma_B} \sim 1/J(0,0)$. In the
opposite limit, $\Gamma_L \ll \Gamma_B$, the dimensionless
coupling $J(\epsilon, \epsilon')$ crosses over from weak
to strong coupling as $|\epsilon'|$ is reduced below 
$k_B T_x \sim \sqrt{\Gamma_L \Gamma_B}$. Anticipating
a relation between $T_K$ and $T_x$, one expects then a
Kondo temperature that is neither exponentially small
in $1/\sqrt{\Gamma_L}$, nor in $1/\sqrt{\Gamma_B}$.

Obviously, the above picture relies heavily on the conjecture
that the Hamiltonian of Eq.~(\ref{H_energy}) flows for
$\epsilon_d = 0$ and $V_B = 0$ to the intermediate-coupling fixed
point of the two-channel Kondo effect, and on intuition borrowed
from the conventional two-channel Kondo Hamiltonian. Although
neither assumption is justified a priori, this tentative
picture is shown below to be surprisingly accurate.

\section{Weak coupling}
\label{sec:weak_coupling}

We begin our discussion with the limit of weak coupling,
$J(0, 0) \ll 1$, for which an analytical treatment is
possible. Specifically, we employ a perturbative scaling
approach based on Anderson's poor-man's
scaling,~\cite{Anderson70} to study two generic cases:
(i) $\epsilon_d = 0$ and $\Gamma_B \ll \Gamma_L$, corresponding
to an impurity level at resonance with the Fermi energy; and
(ii) $|\epsilon_d| \gg \Gamma_B, \Gamma_L$, corresponding to an
impurity level off resonance with the Fermi energy. Throughout
this paper we assume a symmetric rectangular form for the underlying
lead and box density of states, with a single joint bandwidth
$D$.~\cite{comment_on_D}
The latter is taken to be much larger than $|\epsilon_d|, \Gamma_B$,
and $\Gamma_L$, such that $\rho_L^{\rm eff}(\epsilon) = t^2_L
\rho_{L}(\epsilon) |G(\epsilon + i\eta)|^2$ has the
Lorentzian form of Eq.~(\ref{Lorentzian}).

\subsection{Level at resonance with the Fermi energy}
\label{sec:level-at-fermi}

Consider first the case where $\epsilon_d = 0$ and
$\Gamma_B \ll \Gamma_L$. For $\epsilon_d = 0$, the Lorentzian DOS
$\rho_L^{\rm eff}(\epsilon)$ is centered about the Fermi energy.
To proceed with our analytical treatment, it is convenient
to replace $\rho_L^{\rm eff}(\epsilon)$ with a symmetric rectangular DOS
that preserves both the height of $\rho_L^{\rm eff}(\epsilon)$
at the Fermi energy, and its total integrated weight:
\begin{equation}
\rho_L^{\rm eff}(\epsilon) \longrightarrow \frac{1}{\pi \Gamma_L}
       \theta \left ( \frac{\pi\Gamma_L}{2} - |\epsilon| \right ) .
\label{rho_box}
\end{equation}
With this modification, and using the relation
$J(\epsilon, \epsilon') = 2 t_B \sqrt{ \rho_B(\epsilon)
\rho_L^{\rm eff}(\epsilon') }$, the Hamiltonian of
Eq.~(\ref{H_energy}) becomes
\begin{eqnarray}
{\cal H} &=& \sum_{\sigma} \int_{-D_L}^{D_L}
          \epsilon a^{\dagger}_{\epsilon L \sigma}
          a_{\epsilon L \sigma} d\epsilon +
          \sum_{\sigma} \int_{-D}^{D}
          \epsilon a^{\dagger}_{\epsilon B \sigma}
          a_{\epsilon B \sigma} d\epsilon
\nonumber \\
&+& \frac{\tilde{J}_{\perp}}{2} \sum_{\sigma}
          \int_{-D}^{D} \!\! d\epsilon \int_{-D_L}^{D_L} \!\! d\epsilon'
          \!\left \{
                a^{\dagger}_{\epsilon B \sigma} a_{\epsilon' L \sigma}
                S^{+} + {\rm H.c.} \right \}
\nonumber \\
&-& e V_B S_z ,
\label{H_RG_1}
\end{eqnarray}
where $\tilde{J}_{\perp} = (2/\pi) \sqrt{\Gamma_B/\Gamma_L} \ll 1$
is the effective isospin exchange interaction, and
$D_L$ is equal to $\pi \Gamma_L/2$. Hereafter we assume that
$e|V_B| \ll D_L$.

\begin{figure}
\centerline{
\includegraphics[width=85mm]{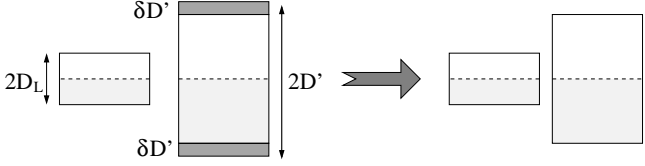}
}\vspace{0pt}
\caption{The basic iterative step in the perturbative scaling
         procedure, for $D > D_L$. Suppose that the isospin-down
         ($B$) bandwidth has already been lowered from its initial
         value $D$ to some smaller value $D > D' > D_L$. The
         bandwidth $D'$ is then further reduced to $D' - \delta D'$
         using poor-man's scaling, while maintaining the isospin-up
         ($L$) bandwidth at $D_L$. This procedure is repeated until
         $D'$ coincides with $D_L$, leaving just a single common
         bandwidth for the two isospin orientations.}
\label{fig:fig2}
\end{figure}

To cope with the different bandwidths in Eq.~(\ref{H_RG_1}),
we proceed with perturbative scaling. Using poor-man's
scaling,~\cite{Anderson70} we successively reduce the larger
bandwidth from $D$ down to $D_L$, mapping thereby the Hamiltonian
of Eq.~(\ref{H_RG_1}) onto an effective low-energy Hamiltonian
with a single joint bandwidth $D_L$. The basic iterative step in
this procedure is illustrated in Fig.~\ref{fig:fig2}.
Suppose that the isospin-down ($B$) bandwidth has already been
lowered from its initial value $D$ to some value $D' = D e^{-l}$,
$0 < l < \ln(D/D_L)$. Further reducing the bandwidth to
$D'(1 - \delta l)$ produces a renormalization to a new interaction
term not present in the original Hamiltonian:
\begin{equation}
\lambda \sum_{\sigma} \int_{-D_L}^{D_L}d\epsilon
         \int_{-D_L}^{D_L} d\epsilon' :\! a^{\dagger}_{\epsilon L \sigma}
                a_{\epsilon' L \sigma}\!: S_z .
\label{H_lambda}
\end{equation}
Here $:\! a^{\dagger}_{\epsilon L \sigma} a_{\epsilon' L \sigma}\!:
= a^{\dagger}_{\epsilon L \sigma} a_{\epsilon' L \sigma}\! -
\theta(-\epsilon) \delta(\epsilon - \epsilon')$ stands for normal
ordering with respect to the filled isospin-down ($L$) Fermi
sea. Explicitly, $\lambda$ renormalizes according to the
scaling equation
\begin{equation}
\frac{d \lambda}{d l} = \frac{1}{2} \tilde{J}_{\perp}^2 ,
\label{lambda_term}
\end{equation}
where $\tilde{J}_{\perp} = (2/\pi) \sqrt{\Gamma_B/\Gamma_L}$
is the bare isospin-exchange coupling in Eq.~(\ref{H_RG_1}).
Indeed, the $\tilde{J}_{\perp}$ interaction term remains
unchanged in Eq.~(\ref{H_RG_1}) throughout this procedure,
apart from the reduced integration range over $\epsilon$. For
a nonzero $\epsilon_d$, there is an additional renormalization
of the ``magnetic'' field $e V_B$, discussed below.

Upon reducing $D$ down to $D_L$, the new coupling $\lambda$
grows from zero to $\frac{1}{2} \tilde{J}_{\perp}^2 \ln (D/D_L)$.
Thus, for $D' = D_L$ one arrives at the effective Hamiltonian
\begin{eqnarray}
{\cal H} &=& \sum_{\sigma} \sum_{\alpha = L, B} \int_{-D_L}^{D_L}
          \epsilon a^{\dagger}_{\epsilon \alpha \sigma}
          a_{\epsilon \alpha \sigma} d\epsilon  - e V_B S_z
\label{H_effective}\\
&+& \frac{\tilde{J}_{\perp}}{2} \sum_{\sigma}
          \int_{-D_L}^{D_L} \!\! d\epsilon \int_{-D_L}^{D_L} \!\! d\epsilon'
          \left \{
                a^{\dagger}_{\epsilon B \sigma} a_{\epsilon' L \sigma}
                S^{+} + {\rm H.c.} \right \}
\nonumber \\
&+& \frac{\tilde{J}_{z}}{2} \sum_{\sigma}
          \int_{-D_L}^{D_L} \!\! d\epsilon \int_{-D_L}^{D_L} \!\! d\epsilon'
          \left \{
                a^{\dagger}_{\epsilon L \sigma} a_{\epsilon' L \sigma}
                - a^{\dagger}_{\epsilon B \sigma} a_{\epsilon' B \sigma}
          \right \} S_z
\nonumber \\
&+& \tilde{V} \sum_{\sigma, \alpha}
          \int_{-D_L}^{D_L} \!\! d\epsilon \int_{-D_L}^{D_L} \!\! d\epsilon'
          :\!a^{\dagger}_{\epsilon \alpha \sigma}
                a_{\epsilon' \alpha \sigma}\!: S_z ,
\nonumber
\end{eqnarray}
where $\tilde{J}_{z} = 2 \tilde{V} = \frac{1}{2}
\tilde{J}_{\perp}^2 \ln (D/D_L)$. Here we have separated the
interaction term of Eq.~(\ref{lambda_term}) into a longitudinal
isospin-exchange interaction $\tilde{J}_{z}$, and a $\tilde{V}$
term.

Apart from the extra $\tilde{V}$ term, Eq.~(\ref{H_effective})
has the form of a conventional two-channel Kondo Hamiltonian
with an anisotropic spin-exchange interaction. As in the case
of a weak tunnel barrier,~\cite{Matveev91} the indices $L$
and $B$ are identified in such a mapping with isospin-up and
isospin-down labels, while the physical spin $\sigma$ serves
as a conserved channel index. Contrary to the case of a weak
tunnel barrier, though, the Hamiltonian of
Eq.~(\ref{H_effective}) contains a longitudinal Kondo
coupling $\tilde{J}_z$, which can be either smaller or larger
than $\tilde{J}_{\perp}$. For $\frac{1}{2} \tilde{J}_{\perp}
\ln (D/D_L) \ll 1$ [i.e., $(1/\pi) \sqrt{\Gamma_B/\Gamma_L}
\ln (D/\Gamma_L) \ll 1$] one has
$\tilde{J}_z \ll \tilde{J}_{\perp}$, while for
$\frac{1}{2} \tilde{J}_{\perp}\ln (D/D_L) \gg 1$ the order is
reversed. It should be emphasized, however, that $\Gamma_L$
(and thus also $\Gamma_B \ll \Gamma_L$) must be exponentially
small in this range in order for $\tilde{J}_z$ to become
comparable to $\tilde{J}_{\perp}$.

It is straightforward to verify using perturbative
renormalization-group (RG) methods that the $\tilde{V}$
term acts much in the same way as
ordinary potential scattering: It is a marginal operator,
that does not affect (at least not to second order) the
Kondo couplings' flow toward strong coupling. For $V_B = 0$,
the Hamiltonian of Eq.~(\ref{H_effective}) thus flows to the
non-Fermi-liquid fixed point of the two-channel Kondo effect,
as in the case of a weak tunnel barrier.~\cite{Matveev91} The
corresponding Kondo temperature can be extracted in turn
from known results for the anisotropic two-channel Kondo model.
In particular, for $\tilde{J}_z \ll \tilde{J}_{\perp}$
this can be done quite elegantly by iterating the
standard RG equations backwards (i.e., {\em increasing}
the bandwidth $D_L$), to obtain a planner two-channel
Kondo Hamiltonian with a bandwidth $D^{\ast} > D_L$ and a
transverse Kondo coupling $\tilde{J}^{\ast}_{\perp}$,
that shares the same $T_K$. To leading order in
$\tilde{J}_{\perp}$ and $\tilde{J}_z/\tilde{J}_{\perp}$
one obtains
\begin{equation}
D^{\ast} = \sqrt{D_L D} \; , \;\;\;\;
\tilde{J}^{\ast}_{\perp} = \tilde{J}_{\perp} =
           \frac{2}{\pi} \sqrt{\frac{\Gamma_B}{\Gamma_L}} .
\label{D^ast}
\end{equation}
Substituting the above parameters into the expression for the
Kondo temperature of the planner two-channel Kondo
Hamiltonian~\cite{T_K_planner_model} yields
\begin{equation}
k_B T_K = \sqrt{\Gamma_B D} \ \exp
           \left [
               -\frac{\pi^2}{4} \sqrt{ \frac{\Gamma_L}{\Gamma_B} }
           \right ].
\label{T_K:weak_coupling}
\end{equation}
Here, as usual, $T_K$ is given up to a factor of order unity,
which depends both on the precise definition of the Kondo
temperature, and on the actual Lorentzian form of
$\rho_L^{\rm eff}(\epsilon)$.

Two comments should be made about
Eqs.~(\ref{D^ast})--(\ref{T_K:weak_coupling}). First, as
seen in Eq.~(\ref{D^ast}), the effective bandwidth for
$\tilde{J}_z \ll \tilde{J}_{\perp}$ is neither $D$ nor $D_L$,
but rather their geometric average, $D^{\ast}$. Therefore, the
effect of the narrow resonance that forms on the level is to
reduce the effective bandwidth in the problem. Second, similar
to the case of a weak tunnel barrier,~\cite{Matveev91} the
exponential dependence in Eq.~(\ref{T_K:weak_coupling}) can be
recast in the form $\exp\!\left [ -\pi^2/2 \sqrt{\cal T} \right]$,
where ${\cal T} = 4\Gamma_B/\Gamma_L$ is the transmission
coefficient through the impurity at the Fermi energy for
$\Gamma_B \ll \Gamma_L$. Hence a resonant level with
$\epsilon_d = 0$ and $\Gamma_B \ll \Gamma_L$ acts similar
to a poorly conducting tunnel barrier with a transmission
coefficient equal to ${\cal T} = 4\Gamma_B/\Gamma_L$.

\subsection{Level off resonance with the Fermi energy}
\label{sec:Level_off_resonance}

Next we consider a level off resonance with the Fermi
energy, namely, $|\epsilon_d| \gg \Gamma_L, \Gamma_B$.
As noted by Gramespacher and Matveev,~\cite{GM00}
tunneling into and out of the quantum box are no longer
equivalent for $\epsilon_d \neq 0$. Depending on the
sign of $\epsilon_d$, this has the effect of either
pushing down or pulling up the charge plateaus. Using
perturbative scaling we show below that a nonzero
$\epsilon_d$ also shifts the position of the degeneracy
point, maintaining the two-channel Kondo effect at the
shifted position of degeneracy point.

To devise a perturbative scaling treatment of the case
$|\epsilon_d| \gg \Gamma_L, \Gamma_B$, one can use either
the original Hamiltonian of Eqs.~(\ref{H_L})--(\ref{H_tun}),
or the equivalent representation of Eq.~(\ref{H_energy}).
In the former representation, one first reduces the
bandwidth from its bare value $D$ to an effective
bandwidth of the order of $|\epsilon_d|$, and then
performs a Schrieffer-Wolff transformation~\cite{SW66}
to eliminate the charge fluctuations on the level.
These two steps enter the representation of
Eq.~(\ref{H_energy}) in a unified fashion through
the energy dependence of the coupling
$J(\epsilon, \epsilon')$, which is sharply peaked
as a function of $\epsilon'$ at $\epsilon_d$.

We have carried out both the perturbative scaling
approach based on the Hamiltonian representation of
Eqs.~(\ref{H_L})--(\ref{H_tun}), and the scheme based on
the Hamiltonian representation of Eq.~(\ref{H_energy}).
Both procedures give the same results to leading order
in $\Gamma_L/|\epsilon_d|$ and $\Gamma_B/|\epsilon_d|$
(note that the Schrieffer-Wolff transformation is designed
to capture only the leading order in these parameters).
For the sake of consistency with the analysis of the
previous subsection, we present below the approach based
on the Hamiltonian representation of Eq.~(\ref{H_energy}).

\begin{figure}[tb]
\centerline{
\vbox{\epsfxsize=85mm \epsfbox{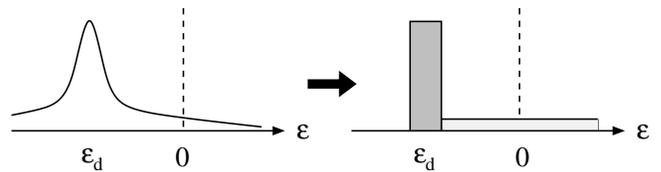}}
}\vspace{0pt}
\caption{Illustration of the density-of-states substitution
         $\rho^{\rm eff}_L(\epsilon) \to
	 \bar{\rho}_{\rm eff}(\epsilon)$, for $\epsilon_d < 0$.
	 For $|\epsilon_d| \gg \Gamma_L$, the Lorentzian
	 $\rho_L^{\rm eff}(\epsilon)$ has two prominent
	 features: a narrow peak of width $\sim \Gamma_L$
	 and weight $w \approx 1$ centered about $\epsilon_d$,
	 and a shallow tail that crosses the Fermi energy.
	 Each of these features is conveniently mimicked
         within $\bar{\rho}_{\rm eff}(\epsilon)$ by a rectangular
	 structure, one narrow and sharp centered about
	 $\epsilon_d$, and the other broad and shallow
	 centered about the Fermi energy.}
\label{fig:fig3}
\end{figure}

Similar to the case of a level at resonance with
the Fermi energy, it is convenient to replace
$\rho_L^{\rm eff}(\epsilon)$ with a simplified
density of states that captures the essential
features of $\rho_L^{\rm eff}(\epsilon)$, and
allows for an analytical treatment of the problem.
Specifically, $\rho_L^{\rm eff}(\epsilon)$ has
two prominent features: a narrow resonance of
width $\sim \Gamma_L$ and weight
$1 - {\cal O}\left( \Gamma_L/|\epsilon_d|\right)$
centered about $\epsilon = \epsilon_d$, and a shallow
tail that crosses the Fermi energy and provides
the low-energy excitations for the development of the
Kondo effect. To mimic these two features, we replace
$\rho_L^{\rm eff}(\epsilon)$ with the double-rectangular
density of states illustrated in Fig.~\ref{fig:fig3}:
\begin{equation}
\bar{\rho}_{\rm eff}(\epsilon) =
           \frac{w}{2 D_L}
           \theta \left ( D_L - |\epsilon - \epsilon_d| \right )
       + \frac{\Gamma_L}{\pi \epsilon_d^2}
                 \theta \left ( D_m - |\epsilon| \right ) .
\label{rho_two_box}
\end{equation}
Here $D_m = |\epsilon_d| - D_L$ is a crude measure of
the extent of the tail that crosses the Fermi energy,
$D_L = \pi \Gamma_L/2$ corresponds to half the width
of the resonance at $\epsilon = \epsilon_d$, and
$w = 1 - 2\Gamma_L D_m/(\pi \epsilon_d^2) \approx 1$
is the effective weight of the resonance.

Obviously, there is some arbitrariness in our choice
of $\bar{\rho}_{\rm eff}(\epsilon)$, which differs in
details from $\rho_L^{\rm eff}(\epsilon)$. In particular,
the extent of the tail that crosses the Fermi energy is
somewhat exaggerated, while the opposite tail (the one
extending away from the Fermi energy) is absent.
Nevertheless, this choice of $\bar{\rho}_{\rm eff}(\epsilon)$
is compatible with the approximations made in the
Schrieffer-Wolff transformation, and gives the correct
results to leading order in $\Gamma_L/|\epsilon_d|$
and $\Gamma_B/|\epsilon_d|$.
Substituting $\bar{\rho}_{\rm eff}(\epsilon)$ in for
$\rho_L^{\rm eff}(\epsilon)$, the Hamiltonian of
Eq.~(\ref{H_energy}) becomes
\begin{eqnarray}
{\cal H} &=& \sum_{\sigma} \int_{-D_p}^{D_p}
          \epsilon a^{\dagger}_{\epsilon L \sigma}
          a_{\epsilon L \sigma} d\epsilon - e V_B S_z 
\label{H_e_d_1} \\
         &+& \sum_{\sigma} \int_{-D}^{D}
          \epsilon a^{\dagger}_{\epsilon B \sigma}
          a_{\epsilon B \sigma} d\epsilon
\nonumber \\
&+& \sqrt{\frac{\Gamma_B}{\pi}} \sum_{\sigma}
          \int_{-D}^{D} \!\! d\epsilon \int_{-D_p}^{D_p} \!\! d\epsilon'
          \!\sqrt{\bar{\rho}_{\rm eff}(\epsilon')}
\nonumber \\
&& \;\;\;\;\;\;\;\;\;\;\;\;\;\;\;\;\;\;\;\;\;\;
   \times \left \{
                a^{\dagger}_{\epsilon B \sigma} a_{\epsilon' L \sigma}
                S^{+} + {\rm H.c.}
          \right \} ,
\nonumber
\end{eqnarray}
where $D_p = |\epsilon_d| + D_L$. Hereafter we assume that
$e|V_B| \ll |\epsilon_d|$.

To treat the Hamiltonian of Eq.~(\ref{H_e_d_1}), we proceed
with perturbative scaling. This is done in two stages. First
the larger bandwidth is successively reduced from its bare
value $D$ down to $D_p$, leaving just a single common
bandwidth for the two conduction seas. This common bandwidth
is subsequently reduced from $D_p$ down to $D_m$.

The first step in the above scheme is similar to the one
implemented in the previous subsection, when treating a level
at resonance with the Fermi energy. There are, however, three
important modifications. First, the new interaction term generated
upon scaling has the form
\begin{equation}
\lambda \sum_{\sigma} \int_{-D_p}^{D_p}d\epsilon
         \int_{-D_p}^{D_p} d\epsilon'
	        \sqrt{\bar{\rho}_{\rm eff}(\epsilon)
		\bar{\rho}_{\rm eff}(\epsilon')}
	        :\! a^{\dagger}_{\epsilon L \sigma}
                a_{\epsilon' L \sigma}\!: S_z ,
\end{equation}
which differs from Eq.~(\ref{H_lambda}) in the extra
square roots of $\bar{\rho}_{\rm eff}$ that enter the
integrand. Second, the coupling $\lambda$ renormalizes
according to the scaling equation
\begin{equation}
\frac{d \lambda}{d l} = \frac{2 \Gamma_B}{\pi} ,
\end{equation}
which likewise differs from Eq.~(\ref{lambda_term}).
Lastly, the voltage $V_B$ is renormalized at $T = 0$
according to
\begin{equation}
\frac{d \tilde{V}_B}{d l} = {\rm sign}(\epsilon_d)
        \frac{2 \Gamma_B}{\pi}
        \frac{1}{1 + e^{l} |\epsilon_d|/D} .
\end{equation}
Here we have distinguished the running parameter
$\tilde{V}_B$ from its bare value $V_B$, and omitted
higher order corrections in $\Gamma_L/|\epsilon_d|$.
Upon reducing the larger bandwidth from $D$ down to
$D_p$, the coupling $\lambda$ thus grows from zero to
$(2\Gamma_B/\pi) \ln(D/D_p)$, while $e\tilde{V}_B$
evolves from $eV_B$ to $eV_B + {\rm sign}(\epsilon_d)(2\Gamma_B/\pi)
\ln\left[ \frac{1}{2} + \frac{D}{2|\epsilon_d|} \right]$.

Since we are interested in $|\epsilon_d| \gg \Gamma_L$,
one can proceed to eliminate all excitations in the
energy range $D_p > \epsilon > D_m$ in one step, by
working with a finite $\delta l = 2 D_L/D_p \ll 1$.
Within the Hamiltonian representation of
Eqs.~(\ref{H_L})--(\ref{H_tun}), this step is
equivalent to carrying out the Schrieffer-Wolff
transformation. At the end of this procedure one
arrives at an effective Hamiltonian of the form
of Eq.~(\ref{H_effective}), where $D_L$ is replaced
with $D_m$, plus two additional potential-scattering
terms:
\begin{equation}
\tilde{V}_{\pm} \sum_{\sigma}
          \int_{-D_m}^{D_m} \!\! d\epsilon
	  \int_{-D_m}^{D_m} \!\! d\epsilon'
          \left \{
                :\!a^{\dagger}_{\epsilon L \sigma}
		a_{\epsilon' L \sigma}\!:
                \pm :\!a^{\dagger}_{\epsilon B \sigma}
		a_{\epsilon' B \sigma} \!:
          \right \} .
\label{V_pm_terms}
\end{equation}
The effective coupling constants entering the resulting
Hamiltonian are quite different, however, from those in
Eq.~(\ref{H_effective}), and are given by
\begin{eqnarray}
&& \tilde{J}_{\perp} = \frac{2}{\pi}
                    \frac{\sqrt{\Gamma_L \Gamma_B}}{|\epsilon_d|} ,
\label{J_perp_e_d} \\
&& \tilde{J}_{z} = \frac{1}{\pi}
                    \frac{\Gamma_B}{|\epsilon_d|} ,
\label{J_z_e_d} \\
&& \tilde{V} = -\frac{1}{2\pi} \frac{\Gamma_B}{|\epsilon_d|} , \\
&& \tilde{V}_{+} = -\frac{1}{4\pi} \frac{\Gamma_B}{|\epsilon_d|}
                    {\rm sign}(\epsilon_d) ,\\
&& \tilde{V}_{-} = \frac{1}{4\pi} \frac{\Gamma_B}{|\epsilon_d|}
                    {\rm sign}(\epsilon_d) ,
\end{eqnarray}
Here we have omitted corrections that are smaller
by factors of $\Gamma_{\alpha}/|\epsilon_d|$ or
$(\Gamma_{\alpha}/\pi|\epsilon_d|)\ln(D/|\epsilon_d|)$
than the leading-order terms. In addition, the voltage
$V_B$ is renormalized at $T = 0$ according to
\begin{equation}
eV_B \to e\tilde{V}_B = eV_B + {\rm sign}(\epsilon_d)
	           \frac{2\Gamma_B}{\pi} \ln
                   \left (
		           1 + \frac{D}{|\epsilon_d|}
		   \right ) .
\label{V_B_e_d}
\end{equation}

It is straightforward to verify using either poor-man's
scaling or bosonization (in combination with a canonical
transformation) that the $V_{\pm}$ terms are marginal,
and do not affect the zero-temperature fixed point of
the Hamiltonian other than through an additional
shift of $\tilde{V}_B$. Specifically, neglecting the
renormalization of $\tilde{J}_z$, Eq.~(\ref{V_B_e_d})
acquires the additional small correction
$e\tilde{V}_B \to e\tilde{V}_B +
D_{p} 16 \ln(2) \left( \tilde{V} \tilde{V}_{+}
+ \tilde{J}_z \tilde{V}_{-}/2 \right)$. Hence the
system continues to undergo the two-channel Kondo
effect for $|\epsilon_d| \gg \Gamma_L, \Gamma_B$,
albeit at a shifted position of the degeneracy
point, approximately given by
\begin{equation}
eV_{\rm 2CK} = -{\rm sign}(\epsilon_d)\frac{2\Gamma_B}{\pi} \ln
                 \left (
		         1 + \frac{D}{|\epsilon_d|}
		 \right ) .
\label{V^star}
\end{equation}

It should be noted, however, that the associated Kondo temperature
is quite sensitive to the ratio of $\Gamma_L$ to $\Gamma_B$, which
fixes the ratio of $\tilde{J}_{\perp}$ to $\tilde{J}_z$ in
Eqs.~(\ref{J_perp_e_d})--(\ref{J_z_e_d}). For
$\Gamma_L \gg \Gamma_B$, one recovers the exponential
form $T_K \propto \exp\!\left [ -\pi^2/2 \sqrt{\cal T} \right]$,
where ${\cal T} = 4 \Gamma_L \Gamma_B/|\epsilon_d|^2$
is the transmission coefficient through the impurity at the
Fermi energy. By contrast, the Kondo temperature depends in
a power-law fashion on $\Gamma_L$, for $\Gamma_B \gg \Gamma_L$:
\begin{equation}
T_K \propto \left (\frac{\Gamma_L}{\Gamma_B}
            \right )^{\pi |\epsilon_d|/2 \Gamma_B} .
\label{T_K_power-law}
\end{equation}
Regardless of the ratio $\Gamma_L/\Gamma_B$, $T_K$
decays exponentially with $|\epsilon_d|$, for
$|\epsilon_d| \gg \Gamma_L, \Gamma_B$.

\section{General coupling}
\label{sec:General_coupling}

Based on perturbative scaling, our treatment thus far was confined
to the weak-coupling regime, $J(0, 0) \ll 1$. We now turn to
a nonperturbative study of all parameter regimes, ranging from
weak to strong coupling. To this end we go back to the
Hamiltonian of Eqs.~(\ref{H_L})--(\ref{H_tun}), and employ
Wilson's numerical renormalization-group (NRG)
method.~\cite{Wilson75} Originally developed for treating
the single-channel Kondo Hamiltonian,~\cite{Wilson75}
this nonperturbative approach was successfully extended to the
Anderson impurity model (both the symmetric~\cite{KWW80a} and
asymmetric~\cite{KWW80b} models), the two-channel Kondo
Hamiltonian,~\cite{Cragg_et_al,PC91} different two-impurity
clusters,~\cite{Jones_et_al,Sakai_et_al,IJW92} and a host of
related zero-dimensional problems. Below we adapt this
approach to the Hamiltonian of Eqs.~(\ref{H_L})--(\ref{H_tun}).

\subsection{The Numerical renormalization group}
\label{sec:NRG}

At the heart of the NRG approach is a logarithmic energy
discretization of the conduction band around the Fermi
energy. The conduction electrons within each energy
interval $[-D\Lambda^{-n}, -D\Lambda^{-n-1}]$ and
$[D\Lambda^{-n-1}, D\Lambda^{-n}]$, $n = 0, 1, 2, \cdots$,
are replaced by a single degree of freedom per spin
orientation, such that all energy intervals contribute
equally to the infra-red divergences which are immanent
in the problem. Here $\Lambda > 1$ is a discretization
parameter, with the full Hamiltonian recovered for
$\Lambda \to 1^+$. Using an appropriate unitary
transformation,~\cite{Wilson75} the conduction band
is mapped onto a semi-infinite chain, with the impurity
coupled to the open end, and the hopping matrix elements
decreasing exponentially along the chain. For the problem
at hand there are two separate bands, one for the lead
and one for the box. Hence four different Wilson shell
operators are required at each point along the chain:
$f^{\dagger}_{\alpha\sigma n}$, where $\alpha = L, B$ labels
the band (lead or box), $\sigma = \uparrow, \downarrow$
is the spin index, and $n = 0, 1, 2, \cdots$ enumerates
the position along the chain. In this manner, the full
Hamiltonian of Eqs.~(\ref{H_L})--(\ref{H_tun}) is recast
as a double limit of a sequence of dimensionless NRG
Hamiltonians:
\begin{equation}
{\cal H} = \lim_{\Lambda \rightarrow 1^+}
	   \lim_{N \rightarrow \infty}
	   \left\{
                 D_{\Lambda} \Lambda^{-(N-1)/2} {\cal H}_{N}
	   \right\} ,
\end{equation}
with $D_{\Lambda}$ equal to $D(1 + \Lambda)^{-1}/2$, and
\begin{eqnarray}
{\cal H}_N &=& \Lambda^{\frac{N-1}{2}}
               \left [
	               \frac{\epsilon_d}{D_{\Lambda}}
		            \sum_{\sigma} \hat{n}_{d \sigma}
                       + \frac{U}{D_{\Lambda}}
		           :\!\hat{n}_{d \uparrow}\!\!:\
                           :\!\hat{n}_{d \downarrow}\!\!:
               \right.  \\
&+&
               \sum_{\sigma}
	       \left \{
	              \tilde{t}_B f^{\dagger}_{B \sigma 0} d_{\sigma} S^{+}
	              + \tilde{t}_L f^{\dagger}_{L \sigma 0} d_{\sigma}
                      + {\rm H.c.}
               \right \}
\nonumber \\
&+&
               \left.
	             \sum_{n = 0}^{N-1}
		     \sum_{\alpha \sigma} \Lambda^{-\frac{n}{2}}
		     \xi_{n \alpha}
		     \left \{
		            f^{\dagger}_{\alpha \sigma n+1}
		            f_{\alpha \sigma n}
	                    + {\rm H.c.}
                     \right \}
               \right ] .
\nonumber
\end{eqnarray}
Here the $\tilde{t}_{\alpha}$'s are related to the hybridization
widths of Eqs.~(\ref{Gamma_L})--(\ref{Gamma_B}) through
$D_{\Lambda}\tilde{t}_{\alpha} = \sqrt{2 \Gamma_{\alpha}D/\pi}$,
while the prefactor $\Lambda^{(N-1)/2}$ guarantees that the
low-lying excitations of ${\cal H}_N$ are of order one for all $N$.

Physically, the shell operators $f^{\dagger}_{\alpha \sigma 0}$
represent the localized states in each band, to which
the impurity level is directly coupled. The
subsequent shell operators $f^{\dagger}_{\alpha \sigma n}$
correspond to wave packets whose spatial
extent about the level increases approximately as $\Lambda^{n/2}$.
All information on the underlying band structure is contained in
the hopping coefficients $\xi_{n \alpha}$,~\cite{comment_on_xi}
which are obtained from appropriate integrals of the density
of states.~\cite{BPH97} In this paper we use a symmetric
rectangular density of states for both the lead and the quantum
box,~\cite{comment_on_dos} for which one has the explicit
expression~\cite{Wilson75}
\begin{equation}
\xi_{n \alpha} = \frac{1-\Lambda^{-(n+1)}}
                 {\sqrt{(1-\Lambda^{-(2n+1)})(1-\Lambda^{-(2n+3)})}} .
\end{equation}

Although the discretized form of the Hamiltonian is exact only in
the limit $\Lambda \to 1^+$, in practice one works with a fixed
value of $\Lambda > 1$. As shown by Wilson,~\cite{Wilson75} the
error introduced by not implementing the limit $\Lambda \to 1^+$
is perturbative and small. Starting with the local Hamiltonian
${\cal H}_0$, the sequence of NRG Hamiltonians are iteratively
diagonalized using the NRG transformation
\begin{equation}
{\cal H}_{N+1} = \sqrt{\lambda}{\cal H}_N
               + \sum_{\alpha, \sigma} \xi_{N \alpha}
	       \left \{
	               f^{\dagger}_{\alpha \sigma N+1}
		       f_{\alpha \sigma N} + {\rm H.c.}
               \right \} .
\label{NRG_iteration}
\end{equation}
At each iteration, four new shell operators are introduced, enlarging
the Hilbert space by a factor of $2^4 = 16$. Since it is numerically
impossible to keep track of such an exponential increase in the
number of basis states, only the lowest $N_s$ eigenstates of
${\cal H}_N$ are retained at each iteration.
These $N_s$ states are used in turn to construct the eigenstates
of ${\cal H}_{N+1}$ using Eq.~(\ref{NRG_iteration}). Thus, two
approximations are involved in the NRG algorithm: discretization
and truncation. Each of these approximations can be systematically
controlled by varying $\Lambda$ and $N_s$.

The eigenstates of ${\cal H}_N$ so obtained,
${\cal H}_{N} |l\rangle_N = E_{l}^{N} |l\rangle_N$, are expected
to faithfully describe the spectrum of the full Hamiltonian
${\cal H}$ on the scale of $D_N = D_{\Lambda} \Lambda^{-(N-1)/2}$.
Hence, they can be used to compute thermodynamic averages
at the temperature $k_B T_N = D_N/\bar{\beta}$,
where $\bar{\beta}$ is a numerical factor of order
unity.~\cite{comment_on_beta_bar} Specifically, the thermodynamic
average of an observable $\hat{\cal O}$ at the temperature $T_N$
is approximated by 
\begin{equation}
\langle \hat{\cal O} \rangle_N = \sum_{l}
                      \frac{e^{-\bar{\beta} E_{l}^{N}}}{Z_N}
		      \ \!_N\langle l| \hat{\cal O}|l\rangle_N ,
\label{NRG_thermodynamic}
\end{equation}
where
\begin{equation}
Z_N = \sum_{l} e^{-\bar{\beta} E_{l}^{N}} .
\label{Z_N}
\end{equation}
In this way, one can compute thermodynamic averages at a
decreasing sequence of temperatures. Note that the effect of
truncation in Eqs.~(\ref{NRG_thermodynamic})--(\ref{Z_N})
can be systematically reduced by increasing the number of
NRG states retained at each iteration.

Apart from the local Hamiltonian ${\cal H}_0$, which
involves the extra $d_{\sigma}$ degrees of freedom, the NRG
formulation of our problem is equivalent to that of the
anisotropic two-channel Kondo Hamiltonian.~\cite{PC91} It
remains so also for an interacting level. Similar to the
anisotropic two-channel Kondo Hamiltonian, the
Hamiltonian of Eqs.~(\ref{H_L})--(\ref{H_tun}) possesses
three underlying symmetries: $SU(2)$ channel symmetry (spin
symmetry), conservation of the total electronic charge, and
conservation of the $z$ component of the total isospin operator:
\begin{equation}
S^{Total}_z = \frac{1}{2} \sum_{k,\sigma}
              \left (
                       c^{\dagger}_{k L \sigma} c_{k L \sigma}
                     - c^{\dagger}_{k B \sigma} c_{k B \sigma}
              \right ) +
	      \frac{1}{2} \sum_{\sigma}
	               d^{\dagger}_{\sigma} d_{\sigma}
            + S_z .
\end{equation}
Each of the NRG Hamiltonians ${\cal H}_{N}$ is block-diagonal in
the conserved quantum numbers, enabling a reduction in the size
of the matrices to be diagonalized. In our code we exploited
only the conservation of the total electronic charge and the $z$
components of the total isospin and physical spin, ignoring the
full $SU(2)$ spin symmetry of the problem. This necessitated
keeping a larger number of NRG states, typically around $N_s = 2300$.

Note that it is straightforward to include a finite on-site
repulsion $U$ within the NRG, as it only enters the local
Hamiltonian ${\cal H}_0$. Apart form modifying the eigen-energies
and eigenstates of ${\cal H}_0$, a finite Coulomb repulsion
$U$ has no effect on the formulation of the NRG. Hence
treatment of an interacting level is computationally
equivalent to that of a noninteracting level.

\begin{figure}[tb]
\centerline{
\vbox{\epsfxsize=75mm \epsfbox{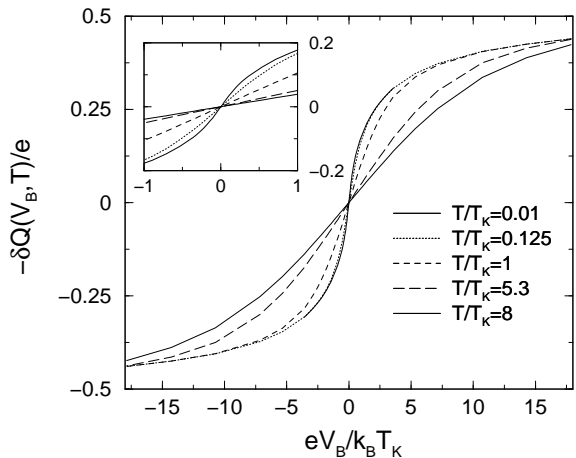}}
}\vspace{0pt}
\caption{Smearing of the charge step at different temperatures.
	 Here $\delta Q = \langle Q \rangle + e(n + \frac{1}{2})$
         is the excess charge inside the box, measured relative
         to the mid point between the two charge plateaus,
         $k_B T_K/D = 0.0014$ is the corresponding two-channel
         Kondo temperature, and $\Gamma_L/D = \Gamma_B/D$ equals
         $0.0039$. The number of NRG states retained is equal to
         2300, and $\Lambda = 2.3$. At temperature $T$, the charge
         step is smeared over a range of
         $eV_B \sim {\rm max}\{k_B T, k_B T_K \}$. The slope of
         the step at $V_B = 0$ continues to steepen with decreasing
         temperature (see inset), diverging logarithmically as
         $T \to 0$.}
\label{fig:fig4}
\end{figure}

\subsection{Level at resonance with the Fermi energy}

\subsubsection{Two-channel Kondo effect}

We begin with a noninteracting level at resonance with the
Fermi energy, i.e., $\epsilon_d = U = 0$. Due to particle-hole
symmetry, the location of the degeneracy point is pinned in
this case at $V_B = 0$.

Figure~\ref{fig:fig4} shows the temperature dependence of
the charge step, for $\Gamma_L/D = \Gamma_B/D = 0.0039$. Here
and throughout the paper we parameterize the charge step by
the excess charge inside the box, measured relative to the
mid point between the two charge plateaus:
\begin{equation}
\delta Q(V_B, T) = \langle Q \rangle
                   + e \left ( n + \frac{1}{2} \right ) .
\end{equation}
For $\epsilon_d = 0$ and equal hybridization widths,
$\Gamma_L = \Gamma_B$, the parameter $J(0, 0)$ in
Eq.~(\ref{J_0_0}) equals $2/\pi \approx 0.64$, which
lies beyond the perturbative scaling analysis of
sec.~\ref{sec:weak_coupling}.

\begin{figure}[tb]
\centerline{
     \includegraphics[width=75mm]{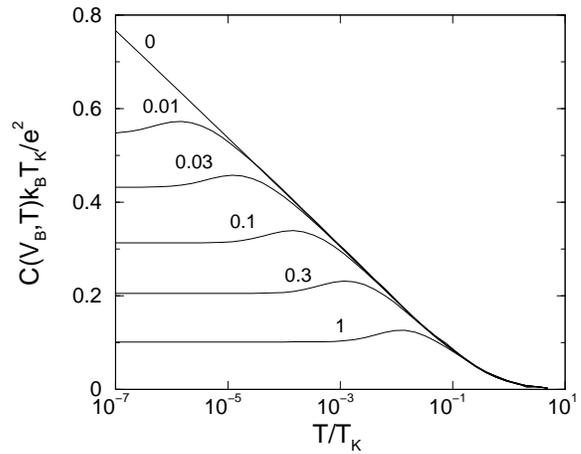}
}\vspace{0pt}
\caption{The capacitance $C(V_B, T)$ versus $T$, for
         $\Gamma_L/D = \Gamma_B/D = 0.0157$ and different values
         of $V_B$. Here $k_B T_K/D = 0.0063$, $\Lambda = 2.3$, and the
         number of NRG states retained is equal to 2300. The deviation
         in voltage from the degeneracy point, $e V_B = 2 E_C \delta N$,
         takes the values $e V_B/ k_B T_K = 0, 0.01, 0.03, 0.1, 0.3$,
         and $1$, according to the labels attached above each
         individual curve.}
\label{fig:fig5}
\end{figure}

As seen in Fig.~\ref{fig:fig4}, the charge step at
temperature $T$ is smeared over a range of
$eV_B \sim {\rm max}\{k_B T, k_B T_K \}$. Here $k_B T_K/D = 0.0014$
is a new low-energy scale, the Kondo temperature, whose
precise definition is given below. In accordance with Matveev's
scenario for a weak tunnel barrier,~\cite{Matveev91} the slope at
$V_B =0$ continues to steepen with decreasing $T$, consistent
with the development of a two-channel Kondo effect. Indeed,
the line shapes in Fig.~\ref{fig:fig4} are quite similar to
those obtained for a weak tunnel barrier using the noncrossing
approximation.~\cite{LSZ01}

The emerges of the two-channel Kondo effect for such
intermediate coupling is confirmed by the capacitance line
shapes, $C(V_B ,T) = - \partial \langle Q \rangle/\partial V_B$,
which are plotted in Fig.~\ref{fig:fig5} for
$\Gamma_L/D = \Gamma_B/D = 0.0157$.~\cite{comment_on_numerics}
At the degeneracy point, $V_B =0$, the capacitance diverges
logarithmically with decreasing temperature, in accordance with
the characteristic $\ln(T)$ divergence of the magnetic
susceptibility in the two-channel Kondo effect.~\cite{CZ98}
For $e |V_B| \ll k_B T_K$, this logarithmic temperature
dependence is cut off about an order of magnitude below
the isospin polarization scale
\begin{equation}
T_{sp} = (e V_B)^2/k_B^2 T_K ,
\label{def_T_sp}
\end{equation}
which governs the crossover from non-Fermi-liquid to
Fermi-liquid behavior (Pauli-like susceptibility). Thus, the
effect of $e V_B$ is identical to that of a magnetic field in the
two-channel Kondo effect.~\cite{CZ98} Indeed, the capacitance
line shapes of Fig.~\ref{fig:fig5} are very similar to the
susceptibility curves obtained by the Bethe ansatz for the
isotropic two-channel Kondo model,~\cite{SS91} with $e V_B$
playing the role of an applied magnetic field.

\begin{figure}[tb]
\centerline{
\includegraphics[width=80mm]{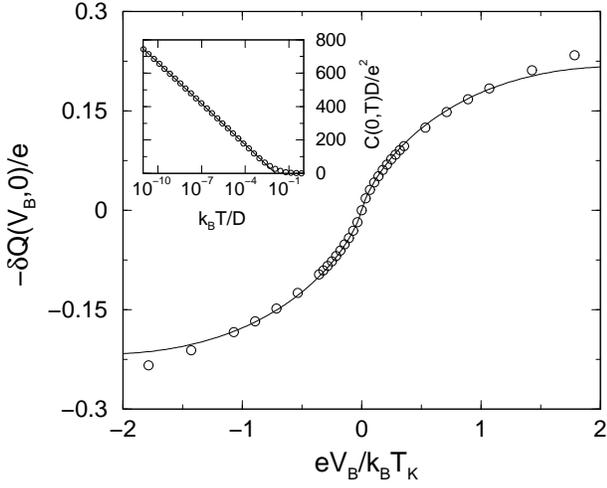}
}\vspace{0pt}
\caption{Zero-temperature smearing of the charge step. Open
	 circles are the calculated NRG points; the full line
	 shows the analytic formula of Eq.~(\ref{q-estimate})
	 with $f = 1.85$. The latter expression for
	 $\delta Q(V_B, 0)$ relies on the logarithmic fit
         $C(0, T) = (e^2/20 k_B T_K)\left[ \ln(T_K/T) + B \right]$,
	 with $k_B T_K/D = 0.0014$ and $B = 2.16$. The quality
	 of this fit for $C(0, T)$ is demonstrated in the inset.
	 Here open circles are the NRG data points, while the
	 full line shows the logarithmic fit. All model and NRG
	 parameters are the same as in Fig.~\ref{fig:fig4}.}
\label{fig:fig6}
\end{figure}

The close resemblance with the Bethe ansatz curves for the
magnetic susceptibility of the two-channel Kondo model provides
us with a precise procedure for defining the two-channel
Kondo temperature $T_K$. Specifically, throughout this paper
we define the two-channel Kondo temperature by the Bethe
ansatz expression for the slope of the $\ln(T)$ diverging
term in the zero-field susceptibility,~\cite{SS89} which
translates in this case to
\begin{equation}
C(0, T) \sim \frac{e^2}{20 k_B T_K}\ln(T_K/T) .
\label{definition_of_Tk}
\end{equation}
Thus, to extract $T_K$ we fitted the capacitance $C(0, T)$ to
the form $(e^2/20 k_B T_K) \left [\ln(T_K/T) + B \right]$,
where $T_K$ and $B$ are fitting parameters. The quality of
our fits is demonstrated in the inset to Fig.~\ref{fig:fig6},
for the same set of model parameters as used in Fig.~\ref{fig:fig4}.

\subsubsection{Shape of the charge step}

As is evident from the curves of Fig.~\ref{fig:fig5}, the
saturated zero-temperature capacitance for $0 < e V_B < k_B T_K$
closely follows the zero-voltage capacitance $C(0, T)$ at the
crossover temperature $T_{sp}$. This suggests that one can
approximate $C(V_B, 0)$ at small voltages by $C(0, T = f T_{sp})$,
where $T_{sp}$ is defined in Eq.~(\ref{def_T_sp}), and $f$ is a
nonuniversal constant of order unity. Using our logarithmic fit
for $C(0, T)$, one can then integrate $C(0, T = f T_{sp})$ with
respect to $V_B$, to obtain the following analytic expression
for the shape of the zero-temperature charge step:
\begin{equation}
\delta Q(V_B, 0) = \frac{e^2 V_B}{20 k_B T_K}
                   \left[
                        2 \ln \left(\frac{|e V_B|}{k_B T_K}\right)
                        - B'
                   \right]
\label{q-estimate}
\end{equation}
with $B' = B + 2 - \ln (f)$.

Figure~\ref{fig:fig6} shows a comparison of Eq.~(\ref{q-estimate})
with the zero-temperature charge step obtained from the NRG, for the
same set of model parameters as used in Fig.~\ref{fig:fig4}. The NRG
data points were obtained by going to a sufficiently low temperature,
such that $\delta Q$ has saturated for each voltage point
displayed at its $T \to 0$ value. As seen in Fig.~\ref{fig:fig6},
Eq.~(\ref{q-estimate}) with $f = 1.85$ well
describes the shape of the zero-temperature charge step up to
$e|V_B| \approx k_B T_K$. For $e|V_B| > k_B T_K$ there are
visible deviations from the NRG data points, which stem from the
breakdown of the logarithmic approximation for $C(V_B, 0)$.

\begin{figure}
\centerline{
    \includegraphics[width=80mm]{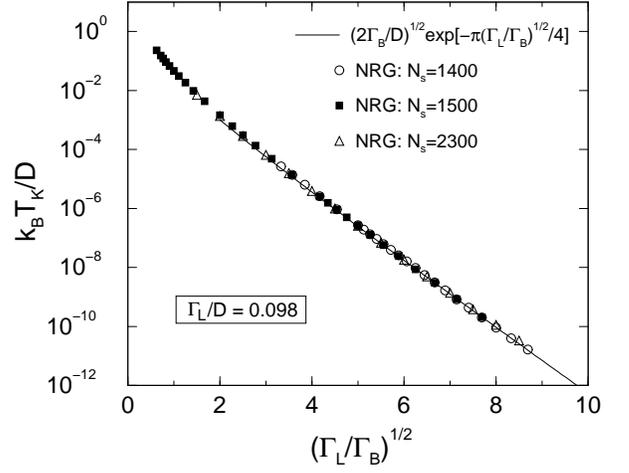}
}\vspace{0pt}
\caption{The Kondo temperature $T_K$ versus $\Gamma_L/\Gamma_B$,
         for $\Gamma_L/D = 0.098$ and $\Gamma_L > \Gamma_B$. Here
         $\Lambda = 2.3$, while the number of NRG states retained
         is varied from $N_s = 1400$ to $N_s = 2300$. For
	 $\Gamma_B \ll \Gamma_L$, the Kondo temperature is
	 well described by the weak-coupling formula of
	 Eq.~(\ref{T_K:weak_coupling}), with a pre-exponential
	 factor equal to $\sqrt{2}$ (solid line).}
\label{fig:fig7}
\end{figure}

\subsubsection{Low-energy scale}

So far, we have focused our attention on intermediate coupling,
$\Gamma_L = \Gamma_B$. Varying $\Gamma_L/\Gamma_B$, we confirmed
that the two-channel Kondo effect persists for all ratios of
$\Gamma_L/\Gamma_B$, ranging from weak ($\Gamma_B \ll \Gamma_L$)
to strong ($\Gamma_B \gg \Gamma_L$) coupling. The main effect of
$\Gamma_L/\Gamma_B$ is to modify the Kondo temperature $T_K$, as
described below.

\begin{figure}
\centerline{
    \includegraphics[width=75mm]{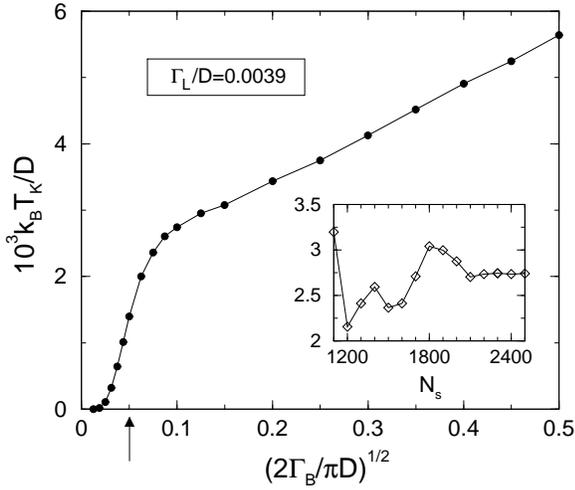}
}\vspace{0pt}
\caption{The Kondo temperature $T_K$ versus $\sqrt{\Gamma_B}$,
         for $\Gamma_L/D = 0.0039$. Here $\Lambda = 2.3$, while
         the number of NRG states retained is equal to 2300. The
         Kondo temperature monotonically increases with $\Gamma_B$,
         initially rapidly for $\Gamma_B < \Gamma_L$, and then
         more moderately for $\Gamma_B > \Gamma_L$. The value
         $\Gamma_B = \Gamma_L$ is marked by an arrow. Inset:
	 Convergence of $T_K$ with the number of NRG states
	 retained $N_s$, for $\Gamma_B = 4\Gamma_L$. Below
	 $N_s = 2000$, finite-size effects are relatively
	 large at this point. Above $N_s = 2100$, the Kondo
	 temperature is converged to within 1\%.}
\label{fig:fig8}
\end{figure}

Figure~\ref{fig:fig7} shows $T_K$ versus $\Gamma_L/\Gamma_B$
in the weak-coupling regime, $\Gamma_B < \Gamma_L$. Here $\Gamma_L/D$
is kept fixed at $0.098$, while $\Gamma_B$ is reduced as to
increase $\Gamma_L/\Gamma_B$. The Kondo temperature was
extracted from the Bethe ansatz expression of
Eq.~(\ref{definition_of_Tk}). For $\Gamma_B \ll \Gamma_L$,
excellent agreement is obtained~\cite{comment_on_A_Lambda} with
the perturbative scaling result of Eq.~(\ref{T_K:weak_coupling}),
up to a pre-exponential factor equal to $\sqrt{2}$. This confirms
the exponential dependence of $T_K$ on $\sqrt{\Gamma_L/\Gamma_B}$.
Deviations from Eq.~(\ref{T_K:weak_coupling}) are observed as
$\Gamma_B$ approaches $\Gamma_L$, signaling the crossover to
intermediate coupling, and the breakdown of perturbative scaling.
Note that the NRG results are practically free of finite-size
effects; only small variations are seen upon going from
$N_s = 1400$ to $N_s = 2300$.

\begin{figure}[tb]
\centerline{
    \includegraphics[width=80mm]{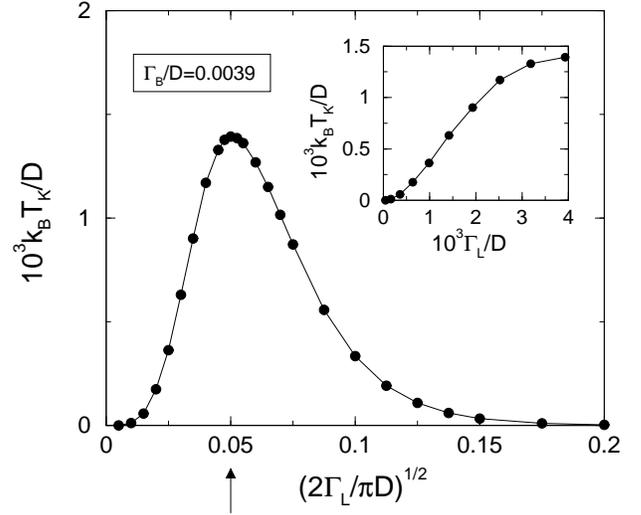}
}\vspace{0pt}
\caption{The Kondo temperature $T_K$ versus $\sqrt{\Gamma_L}$,
         for $\Gamma_B/D = 0.0039$. Here $\Lambda = 2.3$, while
         the number of NRG states retained is equal to 2300.
	 Contrary to Fig.~\ref{fig:fig8}, where $\Gamma_L$ is
         kept fixed, the Kondo temperature increases with
	 increasing $\Gamma_L$ for $\Gamma_L < \Gamma_B$, but
	 then decays exponentially with $\sqrt{\Gamma_L}$ for
         $\Gamma_L > \Gamma_B$. In between, $T_K$ has a
         maximum for $\Gamma_L \approx \Gamma_B$ (the point
	 $\Gamma_B = \Gamma_L$ is marked by an arrow). Inset:
         A plot of $T_K$ versus $\Gamma_L$, for
         $\Gamma_L < \Gamma_B$.}
\label{fig:fig9}
\end{figure}

An extension of Fig.~\ref{fig:fig7} to intermediate and strong
coupling is presented in Fig.~\ref{fig:fig8}. Here a smaller
value of $\Gamma_L/D = 0.0039$ was chosen, as to enable larger
ratios of $\Gamma_B$ to $\Gamma_L$, while maintaining
$\Gamma_B \ll D$. Fixing $\Gamma_L$, the Kondo temperature
monotonically increases as a function of $\Gamma_B$,
revealing two distinct regimes. For weak coupling,
$\Gamma_B \ll \Gamma_L$, one recovers the exponential dependence
of Eq.~(\ref{T_K:weak_coupling}). For $\Gamma_B \gg \Gamma_L$,
this exponential form crosses over to an approximate linear
dependence on $\sqrt{\Gamma_B}$. In the latter regime, $T_K$
is no longer exponentially small in one over the tunneling
matrix elements, as we expand below.

In Fig.~\ref{fig:fig8}, the ratio $\Gamma_L/D = 0.0039$ is kept
fixed and $\Gamma_B$ is varied. Figure~\ref{fig:fig9} shows the
complementary picture, whereby $\Gamma_B/D = 0.0039$ is kept fixed
and $\Gamma_L$ is varied. As is clearly seen from comparison of
Figs.~\ref{fig:fig8} and \ref{fig:fig9}, the two hybridization
widths $\Gamma_L$ and $\Gamma_B$ have inequivalent roles in the
two-channel Kondo effect that develops. In particular, $T_K$
increases monotonically as a function of $\Gamma_B$ for fixed
$\Gamma_L$, but depends nonmonotonically on $\Gamma_L$ when
$\Gamma_B$ is kept fixed. In the latter case, $T_K$ initially
grows with $\Gamma_L$ for $\Gamma_L < \Gamma_B$, but then
decays exponentially with $\sqrt{\Gamma_L}$ for
$\Gamma_L \gg \Gamma_B$. In between, $T_K$ has a characteristic
peak for $\Gamma_L \approx \Gamma_B$.

Experimentally, one can test the predictions of Figs.~\ref{fig:fig8}
and \ref{fig:fig9} by separately tuning $\Gamma_L$ and $\Gamma_B$
using appropriate gate voltages. Perhaps the most natural
parameterization of the combined lead--level--box junction, though,
is in terms of the transmission coefficient through the impurity
at the Fermi energy:
${\cal T} = 4\Gamma_L \Gamma_B/(\Gamma_L + \Gamma_B)^2$. Here we
have set $\epsilon_d = 0$ in the expression for ${\cal T}$, which
depends symmetrically on the hybridization widths $\Gamma_L$ and
$\Gamma_B$. Since $\Gamma_L$ and $\Gamma_B$ have inequivalent
roles in determining $T_K$, it is clear that the Kondo temperature
depends differently on ${\cal T}$, for $\Gamma_L > \Gamma_B$ and
$\Gamma_L < \Gamma_B$.

\begin{figure}[tb]
\centerline{
\vbox{\epsfxsize=80mm \epsfbox{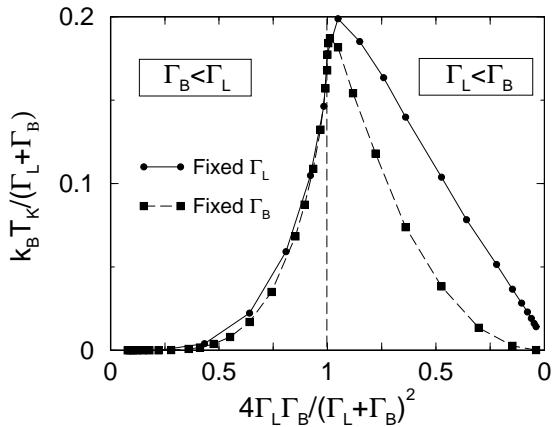}}
}\vspace{0pt}
\caption{The Kondo temperatures of Fig.~\ref{fig:fig8} (where
         $\Gamma_L$ is kept fixed) and Fig.~\ref{fig:fig9}
	 (where $\Gamma_B$ is kept fixed), rescaled by the level
	 broadening $\Gamma_L + \Gamma_B$, and replotted as a
	 function of the transmission through the impurity at
	 the Fermi energy,
         ${\cal T} = 4\Gamma_L\Gamma_B/(\Gamma_L + \Gamma_B)^2$.
         The left-hand side of the graph corresponds to
         $\Gamma_B < \Gamma_L$, while the right-hand side
         corresponds to $\Gamma_L < \Gamma_B$. As a function
         of ${\cal T}$, the ratio $k_B T_K/(\Gamma_L + \Gamma_B)$
         has a peak near perfect transmission. For
         ${\cal T} \approx 1$, the Kondo scale roughly equals
         one fifth of the level broadening.}
\label{fig:fig10}
\end{figure}

In Fig.~\ref{fig:fig10}, we rescaled the Kondo temperatures of
Figs.~\ref{fig:fig8} and \ref{fig:fig9} by the level broadening
$\Gamma_L + \Gamma_B$, and plotted the resulting ratio as a
function of ${\cal T}$. Even after separating the regimes
$\Gamma_L > \Gamma_B$ and $\Gamma_L < \Gamma_B$, the ratio
$k_B T_K/(\Gamma_L + \Gamma_B)$ is not an exclusive function
of ${\cal T}$. Rather, it depends on the
way in which ${\cal T}$ is tuned. Nevertheless, the resulting
${\cal T}$ dependence has several distinct characteristics. Near
perfect transmission, the ratio $k_B T_K/(\Gamma_L + \Gamma_B)$
is greatly enhanced. For $\Gamma_B < \Gamma_L$, this ratio is
peaked at ${\cal T} = 1$, while for $\Gamma_L < \Gamma_B$ the
peak is slightly shifted below ${\cal T} = 1$. Most notably,
$k_B T_K$ roughly equals $0.2(\Gamma_L + \Gamma_B)$ near perfect
transmission, which is many-fold larger than the Kondo temperature
obtained for a tunnel barrier with comparable tunneling matrix
elements.

\begin{figure}[t]
\centerline{
    \includegraphics[width=80mm]{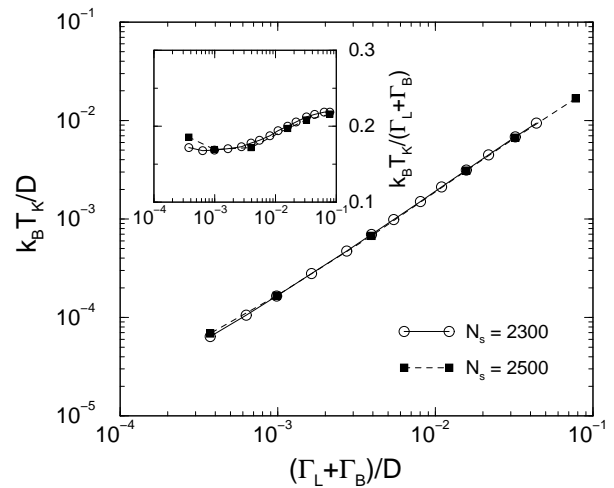}  
}\vspace{0pt}
\caption{$k_B T_K$ versus $\Gamma_L + \Gamma_B$, for $\Gamma_L
         = \Gamma_B$ (i.e., perfect transmission through the
         impurity at the Fermi energy). For over two decades in
         $(\Gamma_L + \Gamma_B)/D$, the Kondo scale $k_B T_K$
	 depends approximately linearly on $\Gamma_L + \Gamma_B$,
	 with a proportionality factor roughly equal to $0.2$
	 (see inset).}
\label{fig:fig11}
\end{figure}

Indeed, while $T_K$ is exponentially small in one over the tunneling
matrix element in the case of a weak tunnel barrier,~\cite{Matveev91}
for $\Gamma_L = \Gamma_B$ it depends approximately linearly on
the level broadening. This important point
is demonstrated in Fig.~\ref{fig:fig11}, where $T_K$ and
$k_B T_K/(\Gamma_L + \Gamma_B)$ are plotted as a function
of $\Gamma_L = \Gamma_B$. For over two decades in
$(\Gamma_L + \Gamma_B)$, the Kondo scale $k_B T_K$ roughly
equals $0.2(\Gamma_L + \Gamma_B)$, establishing the departure
from the familiar exponential form of $T_K$. This dramatic
enhancement of $T_K$ should have important
implications for the observation of the two-channel
Kondo effect in semiconducting devices. Tuning
$\Gamma_L \approx \Gamma_B$ to be smaller than the
charging energy yet notably larger than the level
spacing inside the box, one can possibly realize a Kondo
temperature that significantly exceeds the level spacing,
thereby opening the door to the observation of the
two-channel Kondo effect in semiconducting devices.

It should be noted that finite-size effects
are small in Fig.~\ref{fig:fig11} down to
$(\Gamma_L + \Gamma_B)/D \sim 5\times 10^{-4}$
(compare $N_s = 2300$ with $N_s = 2500$), but rapidly
increase below this value (not shown). Therefore, we are
unable to reliably determine whether or not the approximate
linear dependence of Fig.~\ref{fig:fig11} persists down
to smaller values of $(\Gamma_L + \Gamma_B)/D$.

\subsection{Level off resonance with the Fermi energy}
\label{sec:Level_off_res_nrg}

As discussed in sec.~\ref{sec:Level_off_resonance}, the
position of the degeneracy point is no longer pinned at
$V_B = 0$ for nonzero $\epsilon_d$. Specifically, for
$\epsilon_d \neq 0$ and $k_B T < |\epsilon_d|$, the
average occupation of the level deviates from one-half
per spin orientation. Depending on the sign of $\epsilon_d$,
the level is more available then either for tunneling
into ($\epsilon_d < 0$) or out of ($\epsilon_d > 0$)
the quantum box, which generates a dynamic ``magnetic''
field acting on the isospin. At the degeneracy
point, this dynamic field must be balanced by a nonzero
$V_B$, causing a shift in the position of the degeneracy
point. The resulting low-energy physics at the shifted
degeneracy point is that of the two-channel Kondo
effect, with a reduced Kondo temperature that decays
exponentially with $|\epsilon_d|$, for
$\Gamma_L, \Gamma_B \ll |\epsilon_d|$.

\begin{figure}[tb]
\centerline{
    \includegraphics[width=75mm]{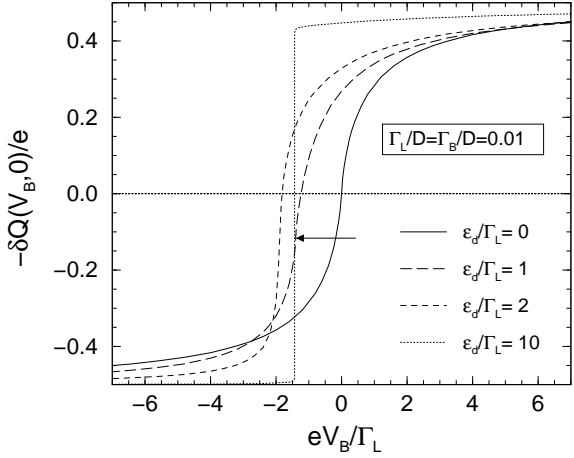}
}\vspace{0pt}
\caption{Evolution of the charge step with increasing
         $\epsilon_d$, for $\Gamma_L/D = \Gamma_B/D = 0.01$.
         Here $\Lambda = 2.3$, while the number of NRG states
         retained is equal to $2300$. With increasing
         $\epsilon_d > 0$, the charge step initially shifts
         toward more negative voltages, before moving back in
         direction of $V_B = 0$. As $\epsilon_d$ exceeds
         $\Gamma_L$, there is a significant narrowing of the
         step separating the two adjacent charge plateaus.
	 For $\epsilon_d = \Gamma_L$, the position of the
         degeneracy point $V_{\rm 2CK}$ is indicated by an
         arrow.}
\label{fig:fig12}
\end{figure}

The above picture was established in sec.~\ref{sec:Level_off_resonance}
using perturbative scaling. We confirmed its validity using
the NRG. In particular, Fig.~\ref{fig:fig12} shows the
zero-temperature charge step [i.e., $\delta Q (V_B, 0)$
versus $V_B$], for $\Gamma_L = \Gamma_B$ and different
values of $\epsilon_d \geq 0$. The corresponding charge
steps for $\epsilon_d \to -\epsilon_d$ are obtained by
simultaneously reflecting the curves of Fig.~\ref{fig:fig12}
about $V_B = 0$ and $\delta Q = 0$.~\cite{comment_on_p-h}
As before, the NRG curves were computed by going to a
sufficiently low temperature, such that $\delta Q$ has
saturated at its $T \to 0$ value for each voltage point
displayed.

With increasing $\epsilon_d > 0$, the charge step initially
shifts toward more negative voltages, before moving back
in direction of $V_B = 0$. The width of the step separating
the two adjacent charge plateaus becomes narrower with increasing
$\epsilon_d$, in accordance with the notion of a decreasing $T_K$.
This narrowing of the step is particularly transparent for larger
values of $\epsilon_d$ (exemplified by $\epsilon_d/\Gamma_L = 10$
in Fig.~\ref{fig:fig12}), when the underlying Kondo temperature
becomes exponentially small in $\epsilon_d/\Gamma_L$. Also
apparent is the development of a slight asymmetry in the
shape of the step. As a function of $V_B$, the approach to
the left charge plateau with $n$ excess electrons in the box
is more rapid than the approach to the right charge plateau
with $n+1$ excess electrons.

One can understand this asymmetry in the shape of the charge
step along the following lines. Consider the limit
$\epsilon_d > (\Gamma_L + \Gamma_B)$, such that the level is
essentially empty. For values of $V_B$ sufficiently removed
to the right of the step, the box is predominantly in the
$n+1$ charge configuration. The weight of the $n$ charge
configuration can then be estimated by simple perturbation theory
with respect to $t_B$, which yields a contribution of order
$t_B^2$. Here we made use of the fact that the empty level is
free to absorb a box electron. By contrast, for values of $V_B$
sufficiently removed to the left of the step, the empty level
has no electron to donate to the box, which blocks any direct
matrix element between the $n$ and $n+1$ charge configurations.
Hence the level must first be charged with a lead electron
before this electron can be passed on to box. This reduces
the weight of the $n+1$ charge configuration by an extra
factor of order $t_L^2$, rendering the $n$ charge
configuration more stable. Therefore, charge fluctuations
are more prominent to the right of the step.

A similar effect, albeit more pronounced due to the larger
values of $\Gamma$ used, is seen in the second-order perturbation
theory of Gramespacher and Matveev,~\cite{GM00} who found that
the charge plateaus are ``pushed down'' for $\epsilon_d > 0$.
Due to the breakdown of perturbation theory, these authors were
unable to access the vicinity of the degeneracy
point. Here we can exploit the NRG to quantitatively trace
the shift in the position of the charge step. Explicitly,
we focus below on the mid-charge point $V^{\ast}$, defined
as the voltage for which $\delta Q(V^{\ast},0) = 0$.

\begin{figure}[tb]
\centerline{
    \includegraphics[width=75mm]{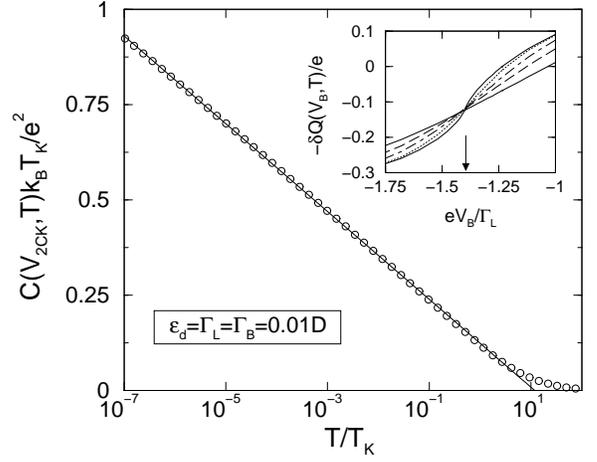}
}\vspace{0pt}
\caption{The capacitance at the shifted degeneracy point,
         for $\epsilon_d = \Gamma_L = \Gamma_B = 0.01 D$.
         Here $k_B T_K/D = 0.0042$, $\Lambda = 2.3$, and
	 $N_s = 2300$. As indicated by the arrow in
	 Fig.~\ref{fig:fig12}, the degeneracy point is
	 slightly shifted away from the position of the
         mid-charge point ($e V_{\rm 2CK}/\Gamma_L = -1.4$
         versus $e V^{\ast}/\Gamma_L = -1.22$). For
         $V = V_{\rm 2CK}$, the capacitance diverges
         logarithmically with $T \to 0$, in accordance with
         the onset of the two-channel Kondo effect. This is
         demonstrated by the solid line, which shows the
         logarithmic fit $C(V_{\rm 2CK}, T)
         = (e^2/20 k_B T_K) \left[ \ln(T_K/T) + B \right]$
         with $k_B T_K/D = 0.0042$ and $B = 2.48$.
	 Inset: Temperature dependence of the charge curve
	 near the degeneracy point. The solid, dotted,
	 dotted-dashed, dashed, and thin-solid lines correspond
	 to $T/T_K = 4\!\times\!10^{-9}, 0.042, 0.15, 0.34$, and
	 $0.78$, respectively. The position of $e V_{\rm 2CK}$
	 is indicated by an arrow.}
\label{fig:fig13}
\end{figure}

An important aspect of the asymmetric line shape for
$\epsilon_d \neq 0$ is a separation between the mid-charge
point $V^{\ast}$ and the degeneracy point $V_{\rm 2CK}$, defined
as the voltage at which two-channel Kondo physics emerges. In the
presence of particle-hole symmetry, $V^{\ast}$ and $V_{\rm 2CK}$
coincide. This is no longer the case away from particle-hole
symmetry, as demonstrated in Figs.~\ref{fig:fig12} and
\ref{fig:fig13} for $\epsilon_d = \Gamma_L = \Gamma_B = 0.01 D$.
Here the degeneracy point is shifted further away from $V_B = 0$
than the mid-charge point ($e V_{\rm 2CK}/\Gamma_L = -1.4$
versus $e V^{\ast}/\Gamma_L = -1.22$). While the mid-charge
point is detected from the condition $\delta Q(V^{\ast},0) = 0$,
the degeneracy point is identified as the point where the
low-temperature charge curves intersect (see inset to
Fig.~\ref{fig:fig13}). As seen in Fig.~\ref{fig:fig13}, the
capacitance diverges logarithmically with decreasing temperature
at $V_B = V_{\rm 2CK}$, confirming the onset of the two-channel
Kondo effect. While clearly not that of a local Fermi liquid, the
finite-size spectrum at $V_B = V_{\rm 2CK}$ deviates from that
of the standard two-channel Kondo Hamiltonian, which is likely
due to the way in which $\epsilon_d \neq 0$ breaks particle-hole
symmetry in the Hamiltonian of Eqs.~(\ref{H_L})--(\ref{H_tun}).
Although $V^{\ast}$ and $V_{\rm 2CK}$ differ for nonzero $\epsilon_d$,
they do closely follow one another. We therefore concentrate
hereafter on the mid-charge point, which is much easier to
trace numerically.

\begin{figure}[tb]
\centerline{
    \includegraphics[width=75mm]{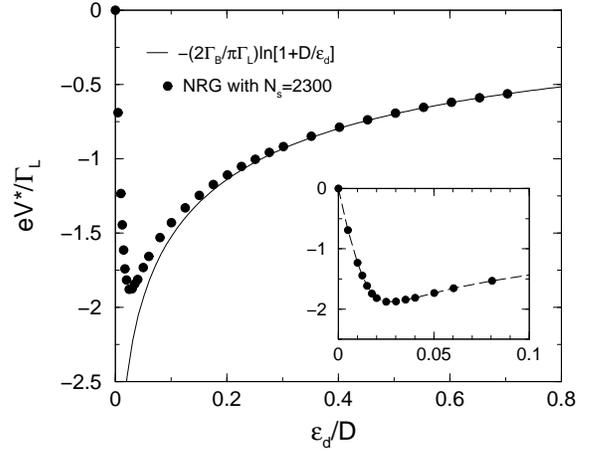}
}\vspace{0pt}
\caption{The shifted position of the mid-charge point
         $V^{\ast}$ versus $\epsilon_d$, for $\Gamma_L/D =
         \Gamma_B/D = 0.01$. Here $\Lambda = 2.3$, while the
         number of NRG states retained is equal to
	 $N_s = 2300$. Solid line: The weak-coupling
	 formula of Eq.~(\ref{V^star}).
	 Inset: An enlarged image of the low-$\epsilon_d$
	 regime. The dashed line is merely a guide for the eye.}
\label{fig:fig14}
\end{figure}

Figure~\ref{fig:fig14} depicts $V^{\ast}$ as a function of
$\epsilon_d > 0$, for $\Gamma_L/D = \Gamma_B/D = 0.01$.
As indicated in Fig.~\ref{fig:fig12}, $V^{\ast}$ has a minimum
at an intermediate energy $\epsilon_d \approx 2.5 \Gamma_L$,
reaching a minimal value of roughly $-2\Gamma_L$. For
$\epsilon_d \gg \Gamma_L$, $V^{\ast}$
gradually increases to zero according to the analytic formula
of Eq.~(\ref{V^star}). The latter expression (which was derived,
strictly speaking, for the degeneracy point $V_{\rm 2CK}$) is
shown for comparison by the solid line in Fig.~\ref{fig:fig14}.
For $\epsilon_d \gg \Gamma_L$, there is good agreement
between the NRG and perturbative scaling.
For $\epsilon_d \gg D$, the position of the mid-charge
point (as well as that of the degeneracy point) once again
approaches $V^{\ast} = 0$, due to the suppression of charge
fluctuations on the level.

Two comments should be made about the predictions of
Figs.~\ref{fig:fig12} and \ref{fig:fig14}. First, one can
experimentally test these predictions by tuning the gate
voltage $V_i$ in the setting of Fig.~\ref{fig:fig1}, which
has the effect of varying $\epsilon_d$. Second, these
predictions are for zero temperature. In general, the
position of the mid-charge point is temperature dependent,
shifting from $V_B = 0$ at $k_B T \gg \epsilon_d$ to
$V_B = V^{\ast}$ at $k_B T \ll \epsilon_d$.

\section{Interacting level}
\label{sec:Interacting_level}

Thus far, our discussion was restricted to a noninteracting
level. However, any realistic setup is bound to include an
on-site repulsion on the level. If the level is realized
by a small quantum dot, this on-site repulsion $U$ will in
fact exceed the charging energy $E_C$, due to the reduced
geometry of the smaller dot. From the standpoint of the
two-channel Kondo effect, the inclusion of a finite $U$
raises several basic questions. Primarily, does the
two-channel Kondo effect persist for an interacting
level? Since an on-site repulsion couples the two spin
orientations on the level, it is not at all clear whether
the physical spin still acts as a passive spectator for the
screening of the charge fluctuations inside the quantum box.
Let us suppose that it does, what is then the effect of a finite
$U$ on the two-channel Kondo temperature? Can one still have a
non-exponential $T_K$? Finally, as is well known, an on-site
repulsion $U$ can itself introduce nontrivial many-body
physics by forming a local magnetic moment on the level.
Can there be a novel interplay between single-channel
screening of the magnetic moment on the level and two-channel
overscreening of the charge fluctuations inside the
quantum box?

In this section we answer these questions, first at the
qualitative level within the framework of perturbative scaling,
and then in a detailed manner using the NRG. Throughout this
section we focus on a symmetric level, $\epsilon_d = - U/2$,
for which the position of the degeneracy point is pinned at
$V_B = 0$ for all $U$. The discussion of the asymmetric case,
$\epsilon_d \neq - U/2$, is left for a forthcoming publication.

\subsection{Weak coupling}
\label{subsec:weak-coupling}

Similar to the case of a noninteracting level, one can exploit
the weak-coupling limit $\Gamma_L \gg \Gamma_B, U$ to gain
some analytic insight as to the effect of a finite $U$.
Starting from the $U = 0$ Hamiltonian of
Eq.~(\ref{H_energy}), the effect of a nonzero $U$ is to
introduce the interaction term
\begin{eqnarray}
{\cal H}_U &=& U \int \!\! d\epsilon_1\!\!
                 \int \!\! d\epsilon_2\!\!
                 \int \!\! d\epsilon_3\!\!
                 \int \!\! d\epsilon_4\
		 W(\epsilon_1, \epsilon_2, \epsilon_3, \epsilon_4)
\label{H_U_energy}\\
&&\;\;\;\;\;\;\;\;\;\;
                    :\!a^{\dagger}_{\epsilon_1 L \uparrow}
                       a_{\epsilon_2 L \uparrow}\!\!:\
                    :\!a^{\dagger}_{\epsilon_3 L \downarrow}
                       a_{\epsilon_4 L \downarrow}\!\!: \ ,
\nonumber
\end{eqnarray}
where
\begin{equation}
W(\epsilon_1, \epsilon_2, \epsilon_3, \epsilon_4) =
              \sqrt{
	              \rho^{\rm eff}_L(\epsilon_1)
	              \rho^{\rm eff}_L(\epsilon_2)
	              \rho^{\rm eff}_L(\epsilon_3)
	              \rho^{\rm eff}_L(\epsilon_4)
                   } .
\end{equation}
Here $\rho^{\rm eff}_L(\epsilon)$ is the effective isospin-up
DOS of Eq.~(\ref{rho_L_eff}), while the normal orderings in
Eq.~(\ref{H_U_energy}) account for the single-particle
energy of the level, $\epsilon_d = -U/2$ [note that we do
not include the latter energy within the Hamiltonian of
Eq.~(\ref{H_energy})]. Replacing $\rho^{\rm eff}_L(\epsilon)$
with the symmetric rectangular DOS of Eq.~(\ref{rho_box}),
the Hamiltonian of Eq.~(\ref{H_energy}) reduces to that
of Eq.~(\ref{H_RG_1}), while the interaction term of
Eq.~(\ref{H_U_energy}) simplifies to
\begin{eqnarray}
{\cal H}_U &=& \frac{U}{(\pi \Gamma_L)^2}
                 \int_{-D_L}^{D_L} \!\! d\epsilon_1\!\!
                 \int_{-D_L}^{D_L} \!\! d\epsilon_2\!\!
                 \int_{-D_L}^{D_L} \!\! d\epsilon_3\!\!
                 \int_{-D_L}^{D_L} \!\! d\epsilon_4\
\nonumber \\
&&\;\;\;\;\;\;\;\;\;\;\;\;
                    :\!a^{\dagger}_{\epsilon_1 L \uparrow}
                       a_{\epsilon_2 L \uparrow}\!\!:\
                    :\!a^{\dagger}_{\epsilon_3 L \downarrow}
                       a_{\epsilon_4 L \downarrow}\!\!: \ .
\label{H_U_box}
\end{eqnarray}

To treat the resulting Hamiltonian, we resort
to perturbative scaling. Similar to the case of a
noninteracting level, this is done in two stages.
First the larger bandwidth of the box is scaled down
from $D$ to $D_L = \pi \Gamma_L/2$, to arrive at an
effective Hamiltonian with a single joint bandwidth $D_L$.
This step is not affected by the presence of a nonzero
$U$, reproducing the Hamiltonian of Eq.~(\ref{H_effective})
with the extra interaction term of Eq.~(\ref{H_U_box}).
Once a single bandwidth is reached, one can proceed
with conventional RG. Successively reducing the single
bandwidth using poor-man's scaling, one notes the
following properties:
\begin{itemize}
\item
      Of the two isospin Kondo couplings $\tilde{J}_{\perp}$
      and $\tilde{J}_z$, only the renormalization of
      $\tilde{J}_z$ is directly affected by the interaction
      term of Eq.~(\ref{H_U_box}).
\item
     All interaction terms generated under the RG conserve the
     physical spin, containing an equal number of creation
     and annihilation operators for each spin orientation.
     In particular, aside from the renormalization to
     $\tilde{J}_{\perp}$ and $\tilde{J}_z$, the
     isospin-exchange interaction retains its original
     two-channel form.
\item
    The Coulomb interaction $U$ has a scaling dimension
    of two, and is formally irrelevant. The same is true
    of all higher order interaction terms generated
    under the RG, involving four fermion operators or more.
\end{itemize}
Thus, all higher order interaction terms tend to diminish
as the bandwidth is reduced, leaving only the three
interaction terms Eq.~(\ref{H_effective}):
$\tilde{J}_{\perp}$, $\tilde{J}_z$, and $\tilde{V}$. We therefore
conclude that the two-channel Kondo effect is robust for
$\Gamma_L \gg \Gamma_B$ against the inclusion of a weak
on-site repulsion $U$. The latter has the effect of only
weakly modifying the Kondo temperature $T_K$.

\subsection{Large on-site repulsion}
\label{subsec:large-u}

Another limit largely tractable by analytical means is that of a
large repulsion on the level, $U \gg \Gamma_L, \Gamma_B, e|V_B|$.
In this limit, a stable local moment is formed on the level as
the temperature is reduced below $U/2$. Therefore, one can resort
to a generalized Schrieffer-Wolff transformation,~\cite{SW66}
to eliminate the charge fluctuations on the level. This produces
an effective low-energy Hamiltonian, which can be treated in turn
using perturbative RG.

To implement this strategy, we go back to the Hamiltonian of
Eqs.~(\ref{H_L})--(\ref{H_tun}) and (\ref{Hubbard}), and
consider the limit of a large repulsion on the level, $U > D$.
Carrying out a Schrieffer-Wolff-type transformation,
a host of spin-exchange $\otimes$ isospin-exchange interactions
are generated. These are conveniently expressed in terms of two
sets of operators, acting as spin-exchange and potential scattering
from the standpoint of the physical spin:
\begin{equation}
\hat{O}_{\mu \nu}^s = \frac{1}{4} S_{\mu}\!
             \sum_{\alpha, \beta = L, B}\!
	          \sigma^{\nu}_{\alpha \beta}
             \sum_{\sigma, \sigma'}\!
	          \vec{\tau} \cdot \vec{\sigma}_{\sigma \sigma'}
             \sum_{k, k'}
	          :\! c^{\dagger}_{k\sigma\alpha} c_{k'\sigma'\beta}\!: ,
\end{equation}
\begin{equation}
\hat{O}_{\mu \nu}^{c} = \frac{1}{4} S_{\mu}\!
             \sum_{\alpha, \beta = L, B}\!
	          \sigma^{\nu}_{\alpha \beta}
             \sum_{k, k', \sigma}
	          :\! c^{\dagger}_{k\sigma\alpha} c_{k'\sigma\beta}\!:
\end{equation}
($\mu, \nu = 0, x, y, z$). Here we have represented the magnetic
moment on the level by the spin-$\frac{1}{2}$ operator
\begin{equation}
\vec{\tau} = \frac{1}{2} \sum_{\sigma, \sigma'}
                 \vec{\sigma}_{\sigma \sigma'}
		        d^{\dagger}_{\sigma} d_{\sigma'}
\label{tau-def}
\end{equation}
($\vec{\sigma}$ being the Pauli matrices), and identified the isospin
indices $L$ and $B$ with isospin-up and isospin-down labels. Normal
ordering of the conduction-electron operators is with respect to the
filled Fermi seas of the lead and the box. We further use the notation
by which $S_0$ is the unity operator in the isospin space, and
$\sigma_0$ is the $2\!\times\!2$ unity matrix. In terms of the above
operators, the resulting Hamiltonian is compactly expressed as
\begin{eqnarray}
{\cal H}_{\rm SW} &=& \sum_{k,\sigma, \alpha} \epsilon_{k \alpha}
             c^{\dagger}_{k \alpha \sigma} c_{k \alpha \sigma}
             - e V_B S_z
\label{H_SW} \\
&& + {\cal J}_{0} \hat{O}^s_{00} + {\cal J}_{z} \hat{O}^s_{0z}
   + {\cal J}_{\perp}
                         \left[
	                       \hat{O}^s_{xx} +
			       \hat{O}^s_{yy}
			 \right]
\nonumber \\
&& + V_{z} \hat{O}^c_{zz} + V_{0} \hat{O}^c_{z0} ,
\nonumber
\end{eqnarray}
where
\begin{eqnarray}
&& {\cal J}_{0} = \frac{8 t_L^2 + 4 t_B^2}{U} ,
\label{J_0_bare} \\
&& {\cal J}_{z} = \frac{8 t_L^2 - 4 t_B^2}{U} ,\\
&& {\cal J}_{\perp} = 16 \frac{t_L t_B}{U} ,\\
&& V_{z} = 4\frac{t_B^2}{U} ,\\
&& V_{0} = -4\frac{t_B^2}{U} .
\label{V_0_bare}
\end{eqnarray}

Physically, the spin-exchange interactions $\hat{O}_{00}^s$ and
$\hat{O}_{0z}^s$ do not scatter electrons between the lead and
the box, conserving thereby the isospin component $S_z$. By
contrast, the spin-exchange interaction
$\hat{O}_{xx}^s + \hat{O}_{yy}^s$ involves hopping between the
lead and the box. This separation of processes is analogous
to that of intra-lead coupling and inter-lead coupling in the
conventional two-lead Kondo problem. The additional
$\hat{O}_{zz}^c$ and $\hat{O}_{zo}^c$ terms are equivalent in
turn to the $\tilde{J}_z$ and $\tilde{V}$ interactions in the
Hamiltonian of Eq.~(\ref{H_effective}).

Proceeding with perturbative RG, no new interaction terms
are generated to second order as the bandwidth is reduced.
Converting to the dimensionless couplings
$\tilde{\cal J}_{\mu} = \rho {\cal J}_{\mu}$ and
$\tilde{V}_{\nu} = \rho V_{\nu}$ with $\mu = 0, z, \perp$
and $\nu = 0, z$ (for simplicity we assume a common density
of states $\rho$ for the lead and the box), the dimensionless
couplings evolve according to the RG equations
\begin{eqnarray}
&& \frac{d \tilde{\cal J}_0}{dl} = \frac{1}{2}
           \left[
                  \tilde{\cal J}_0^2 + \tilde{\cal J}_z^2
           \right]
	   + \frac{1}{4} \tilde{\cal J}_{\perp}^2 ,
\label{RG_J_0} \\
&& \frac{d \tilde{\cal J}_z}{dl} =
           \tilde{\cal J}_0 \tilde{\cal J}_z , \\
&& \frac{d \tilde{\cal J}_{\perp}}{dl} =
           \tilde{\cal J}_0 \tilde{\cal J}_{\perp}
	   + \frac{1}{2} \tilde{V}_z \tilde{\cal J}_{\perp} , \\
&& \frac{d \tilde{V}_z}{dl} =
           \frac{3}{8} \tilde{\cal J}_{\perp}^2 ,\\
&& \frac{d \tilde{V}_0}{dl} = 0 .
\label{RG_V_0}
\end{eqnarray}
Here $l$ is equal to $\ln (D/D')$, where $D'$ is the running
bandwidth and $D$ is the bare bandwidth.

\begin{figure}
\centerline{
\vbox{\epsfxsize=75mm \epsfbox{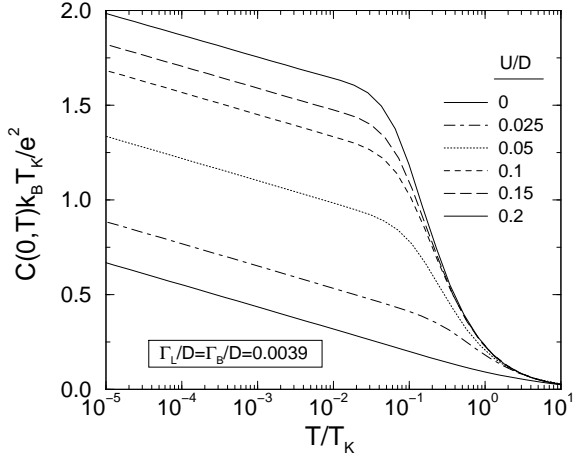}}
}\vspace{0pt}
\caption{The scaled capacitance at the degeneracy point,
         $C(0, T) k_B T_K/e^2$, versus $T$, for different values
	 of the on-site repulsion $U$. Here $\Gamma_L/D = \Gamma_B/D
         = 0.0039$, $\Lambda = 2.3$, and the number of NRG
         states retained is equal to $N_s = 2300$. With increasing
         $U$, a sharp shoulder develops in $C(0, T)$ just
         prior to the onset of the characteristic two-channel
         logarithmic temperature dependence,
         $C(0, T) \sim (e^2/20 k_B T_K) \ln(T_k/T)$. Concurrent
         with the development of the sharp shoulder, the
	 onset of the $\ln(T)$ temperature dependence is
	 pushed down in temperature from
	 $T/T_K \sim 0.2$ for $U \ll \Gamma_L$, to
	 $T/T_K \sim 0.02$ for $U \gg \Gamma_L$.}
\label{fig:fig15}
\end{figure}

Although Eqs.~(\ref{RG_J_0})--(\ref{RG_V_0}) have no simple
analytic solution, one can read off the essential physics
from the structure of these equations and the initial
conditions of Eqs.~(\ref{J_0_bare})--(\ref{V_0_bare}).
Primarily, the system flows toward strong coupling for any
ratio of $\Gamma_L$ to $\Gamma_B$, indicating the emergence
of a Kondo effect for any $\Gamma_L, \Gamma_B \ll U$. For
$\Gamma_L \sim \Gamma_B$, the coupling $\tilde{\cal J}_{\perp}$
is the largest throughout the RG flow, and is the first
to become of order unity. Hence the magnetic moment $\vec{\tau}$
and the isospin $\vec{S}$ are simultaneously quenched.
By contrast, for either $\Gamma_L \gg \Gamma_B$ or
$\Gamma_B \gg \Gamma_L$, the couplings $\tilde{\cal J}_0$
and $\tilde{\cal J}_z$ both grow to order unity at some
characteristic temperature $T^{\ast}$, while
$\tilde{\cal J}_{\perp}$ and $\tilde{V}_z$ remain small
at this temperature. In this case, the isospin $\vec{S}$
remains essentially free when the magnetic moment
$\vec{\tau}$ is screened either by the lead electrons
(for $\Gamma_L \gg \Gamma_B$) or by the box electrons
(for $\Gamma_B \gg \Gamma_L$). A second-stage quenching
of $\vec{S}$ is expected at some lower temperature, yet
this low-temperature regime lies beyond the range of
validity of Eqs.~(\ref{RG_J_0})--(\ref{RG_V_0}). Below
we confirm the two-stage quenching of $\vec{\tau}$ and
$\vec{S}$ using the NRG.

It should be emphasized that, regardless of the ratio
$\Gamma_L/\Gamma_B$, the above analysis is insufficient
for determining the precise nature of the low-temperature
fixed point, whether the isospin moment is exactly screened
or overscreened. As we show below using the NRG, the low-temperature
fixed point is indeed that of an overscreened isospin moment,
for all values of $\Gamma_L, \Gamma_B \neq 0$.

\subsection{General on-site repulsion}

Following the analytic analysis presented above, we now turn
to a systematic study of all coupling regimes using the NRG.
As emphasized in sec.~\ref{sec:NRG}, it is straightforward
to incorporate a nonzero repulsion $U$ within the NRG, as
it only enters the local Hamiltonian ${\cal H}_0$. The
computational effort for treating a nonzero $U$ remains the
same as that for a noninteracting level.

\begin{figure}
\centerline{
\vbox{\epsfxsize=75mm \epsfbox{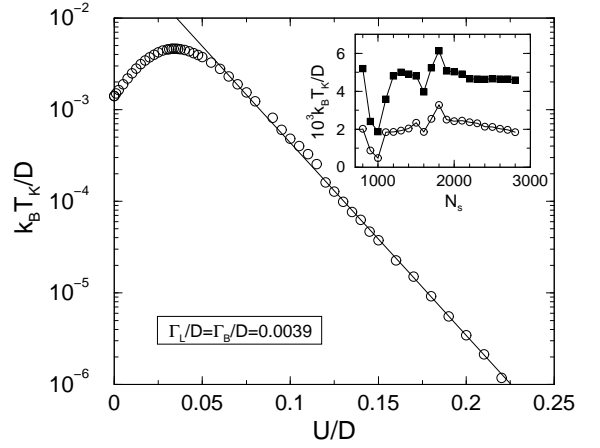}}
}\vspace{0pt}
\caption{The two-channel Kondo temperature $T_K$ versus
         $U$, for $\Gamma_L/D = \Gamma_B/D = 0.0039$.
         Here $\Lambda = 2.3$, while the number of NRG states
         retained is equal to $N_s = 2300$. With increasing
	 $U$, the Kondo temperature initially grows, reaching
         a maximum for $U/\Gamma_L \approx 9$, and then decays.
         For $U \gg \Gamma_L + \Gamma_B$, when the level is
         in the local-moment regime, $T_K$ can be fitted by the
	 exponential form $k_B T_K/D = A \exp( -B U/\Gamma_L)$,
	 where $A = 0.055$ and $B = 0.19$ (solid line). Inset:
         Convergence of $T_K$ with the number of NRG states
         retained, for $U/\Gamma_L = 8.9$ (filled squares)
         and $U/\Gamma_L = 16.5$ (empty circles).
         For $U/\Gamma_L = 8.9$, $T_K$ is converged to
         within 2\% above $2200$ states. For $U/\Gamma_L
         = 16.5$, $T_K$ still varies by some 25\% in going
         from $N_s = 2300$ to $N_s = 2800$.}
\label{fig:fig16}
\end{figure}

Analyzing the finite-size spectra generated by the NRG,
we find that the low-temperature fixed point remains
that of the two-channel Kondo effect, for all values
of $U, \Gamma_L$, and $\Gamma_B$ explored. Accordingly,
the capacitance $C(0, T)$ diverges logarithmically with
$T \to 0$ for all values of $U$, as demonstrated in Fig.~\ref{fig:fig15}
for $\Gamma_L = \Gamma_B$. The effect of a finite $U$
in Fig.~\ref{fig:fig15} is most clearly seen in the
crossover regime, prior to the onset of the characteristic
two-channel logarithmic temperature dependence of the
capacitance. With increasing $U$, a sharp shoulder
develops in $C(0, T)$ just above the onset of the
$\ln(T)$ temperature dependence, which in turn is
pushed down in temperature from $T/T_K \sim 0.2$ for
$U \ll \Gamma_L$ to $T/T_K \sim 0.02$ for
$U \gg \Gamma_L$. Here $T_K$ is the two-channel
Kondo temperature, extracted from a logarithmic
fit to Eq.~(\ref{definition_of_Tk}). As we show below,
the sharp shoulder that develops in $C(0, T)$ is a
signature of the simultaneous quenching of the isospin
$\vec{S}$ and the local magnetic moment $\vec{\tau}$
that forms on the level for a large $U$. It is lost for
$\Gamma_L \gg \Gamma_B$, when $\vec{\tau}$ is quenched well
ahead of the isospin $\vec{S}$ (see Fig.~\ref{fig:fig19}).

Figure~\ref{fig:fig16} shows the two-channel Kondo temperature
$T_K$ versus $U$, for $\Gamma_L/D = \Gamma_B/D = 0.0039$. Quite
surprisingly, the Kondo temperature initially grows with
increasing $U$, reaching a maximum for $U/\Gamma_L \approx 9$.
This regime of enhanced $T_K$ is neither covered by the
weak-coupling analysis of sec.~\ref{subsec:weak-coupling},
nor by the large-$U$ treatment of sec.~\ref{subsec:large-u}.
Note, however, that despite the three-fold enhancement of
$T_K$ as compared to the $U = 0$ case, the logarithmic
temperature dependence of the capacitance sets in at a
lower temperature for $U/\Gamma_L \approx 9$ than for
$U = 0$. For a large on-site repulsion,
$U \gg \Gamma_L + \Gamma_B$, the Kondo temperature
decays exponentially with $U$.

\begin{figure}
\centerline{
\vbox{\epsfxsize=75mm \epsfbox{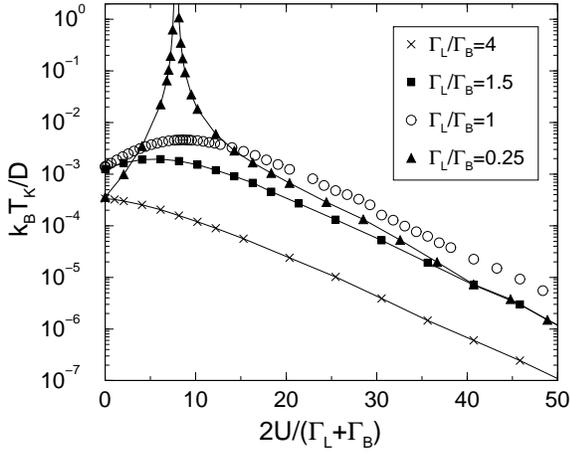}}
}\vspace{0pt}
\caption{The two-channel Kondo temperature $T_K$ versus
         $U$, for $\Gamma_B/D = 0.0039$ and different
	 ratios of $\Gamma_L$ to $\Gamma_B$. Here
	 $\Lambda = 2.3$, while the number of NRG states
         retained is equal to $N_s = 2300$. With
	 increasing $\Gamma_L/\Gamma_B > 1$, the
	 peak in $T_K$ as a function of $U$ becomes
	 shallower, until it disappears. In particular,
	 no peak is left for $\Gamma_L/\Gamma_B = 4$.
	 By contrast, the peak sharpens with decreasing
	 $\Gamma_L/\Gamma_B < 1$.}
\label{fig:fig17}
\end{figure}

Repeating the calculation of $T_K$ versus $U$ for different
ratios of $\Gamma_L$ to $\Gamma_B$, we find a qualitative
difference between $\Gamma_L > \Gamma_B$ and
$\Gamma_B > \Gamma_L$. As seen in Fig.~\ref{fig:fig17},
the peak in $T_K$ as a function of $U$ becomes shallower
with increasing $\Gamma_L/\Gamma_B > 1$, until it
disappears. For $\Gamma_L/\Gamma_B = 4$, for example,
there is no peak left. The slope of $T_K$ versus $U$ at
$U = 0$ also becomes shallower as $\Gamma_L/\Gamma_B$ is
increased, in agreement with the perturbative scaling
analysis of sec.~\ref{subsec:weak-coupling}. The latter
predicts a weak $U$ dependence of the Kondo temperature
for $U, \Gamma_B \ll \Gamma_L$.

In contrast, the peak in $T_K$ versus $U$ becomes sharper
and higher as $\Gamma_L/\Gamma_B < 1$ is decreased. For
$\Gamma_L/\Gamma_B = 1/4$, $T_K$ actually exceeds the
conduction-electron bandwidth as one approaches the peak
position. In this range of $U$, there is a clear separation
(over three orders of magnitude) between the low-temperature
scale $T_K$, extracted from the slope of the $\ln(T)$ component
of $C(0, T)$, and the thermodynamic crossover scale $T_0$,
below which the $\ln(T)$ temperature dependence sets in.
In fact, it becomes increasingly difficult to extract a meaningful
$T_K$ from the slope of the $\ln(T)$ diverging term as one
approaches the peak position, possibly signaling the breakdown
of the two-channel Kondo effect at some critical $U$. We
emphasize, however, that the low-temperature fixed point
revealed by the NRG remains that of the two-channel Kondo
effect, for all values of $U$ explored.

\begin{figure}
\centerline{
\vbox{\epsfxsize=85mm \epsfbox{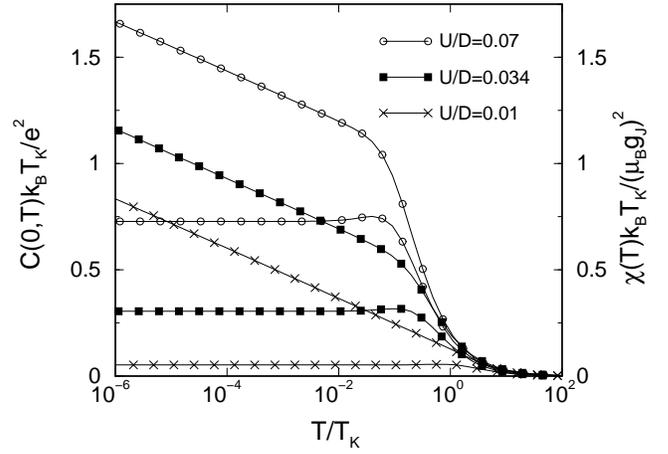}}
}\vspace{0pt}
\caption{The capacitance (unsaturated curves) versus the magnetic
         susceptibility of the level (saturated curves),
         for $\Gamma_L/D = \Gamma_B/D = 0.0039$ and different
	 values of the on-site repulsion $U$. Here $\Lambda = 2.3$,
	 while the number of NRG states retained is equal to
	 $N_s = 2300$. Both the capacitance and the magnetic
	 susceptibility are scaled with the two-channel Kondo
	 temperature $T_K$, extracted from the slope of the
	 $\ln(T)$ diverging term in the capacitance. Explicitly,
	 $k_B T_K/D$ is equal to $0.0025$, $0.0047$, and $0.0019$
	 for $U/D = 0.01, 0.034$, and $0.07$, respectively.
	 For $U/D = 0.01$, the level is on the boarder line
	 between the mixed-valent and local-moment regimes.
	 As $U$ is increased, a stable local moment forms on
	 the level. In this regime, the onset of the $\ln(T)$
	 temperature dependence of the capacitance and the
	 saturation of the magnetic susceptibility take place
	 roughly at the same crossover temperature.}
\label{fig:fig18}
\end{figure}

Up to this point, we focused our discussion on the overscreening
of the isospin $\vec{S}$, as probed by the capacitance.
However, for a large $U$, a local magnetic moment is
formed on the level. To clarify the interplay between the
screening of the magnetic moment and the overscreening of
the isospin $\vec{S}$, we compare in Figs.~\ref{fig:fig18}
and \ref{fig:fig19} the capacitance (i.e., the isospin
susceptibility) with the magnetic susceptibility of
the level. To this end, we augment the Hamiltonian of
Eqs.~(\ref{H_L})--(\ref{H_tun}) and (\ref{Hubbard}) with
the local magnetic field
\begin{equation}
{\cal H}_{mag} = -\mu_B g_J H \tau_z ,
\end{equation}
where $\tau_z$ is the $z$ component of the spin operator
defined in Eq.~(\ref{tau-def}). The magnetic susceptibility
of the level is given in turn by the derivative
\begin{equation}
\chi(T) = \mu_B g_J
          \frac{\partial \langle \tau_z \rangle}{\partial H} ,
\end{equation}
evaluated at zero field.

Figure~\ref{fig:fig18} shows our results for
$\Gamma_L/D = \Gamma_B/D = 0.0039$. In the local-moment regime,
the magnetic susceptibility saturates at low temperatures,
much in the same way as it does in the conventional
one-channel Kondo effect. In accordance with the large-$U$
analysis of sec.~\ref{subsec:large-u}, the saturation of
$\chi(T)$ occurs at the same temperature range as the
onset of the $\ln(T)$ temperature dependence of $C(0, T)$,
confirming the simultaneous quenching of the spin $\vec{\tau}$
and the isospin $\vec{S}$. Indeed, the zero-temperature
susceptibility $\chi(0)$ is of the order of $(\mu_B g_J)^2/T_K$,
indicating that the same Kondo scale $T_K$ underlies both
the screening of $\vec{\tau}$ and the overscreening of
$\vec{S}$.

\begin{figure}
\centerline{
\vbox{\epsfxsize=85mm \epsfbox{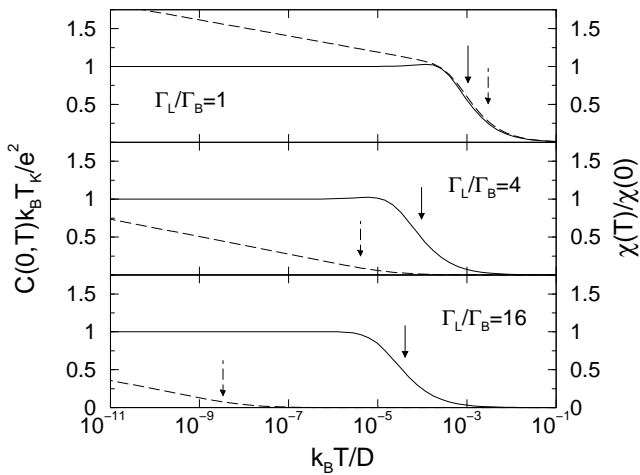}}
}\vspace{0pt}
\caption{The capacitance (dashed lines) versus the magnetic
         susceptibility of the level (solid lines), for $U/D = 0.3$,
	 $\Gamma_L/D = 0.0157$, and different ratios of
	 $\Gamma_L$ to $\Gamma_B$. Here $\Lambda = 2.3$,
	 while the number of NRG states retained is equal to
         $N_s = 2300$. The capacitance is scaled with the
	 two-channel Kondo temperature $T_K$, extracted from
	 the slope of its $\ln(T)$ diverging term. The magnetic
	 susceptibility is scaled with its zero-temperature
	 value, which defines yet another Kondo scale
	 $k_B T_{\rm 1ch} = (\mu_B g)^2/4 \chi(0)$. For
	 $\Gamma_L = \Gamma_B$, the magnetic susceptibility
	 saturates at the same temperature range where the
	 $\ln(T)$ temperature dependence of the capacitance
	 sets in. As the ratio $\Gamma_L/\Gamma_B$ is increased,
	 the impurity moment is quenched well ahead of the
         isospin $\vec{S}$. Accordingly, a large separation
	 builds up between the associated one-channel and
	 two-channel Kondo scales $k_B T_{\rm 1ch}$ and $k_B T_K$
	 (marked by the full and dashed arrows, respectively).}
\label{fig:fig19}
\end{figure}

As the ratio $\Gamma_L/\Gamma_B$ is increased, see
Fig.~\ref{fig:fig19}, two distinct Kondo scales emerge, one
associated with the single-channel screening of $\vec{\tau}$,
and the other associated with the two-channel overscreening of
$\vec{S}$. As before, the two-channel Kondo temperature $T_K$
is extracted from a logarithmic fit of the capacitance to
Eq.~(\ref{definition_of_Tk}), while the one-channel Kondo
temperature is defined from the zero-temperature magnetic
susceptibility:
\begin{equation}
k_B T_{\rm 1ch} = \frac{(\mu_B g)^2}{4 \chi(0)} .
\end{equation}

For $\Gamma_L = \Gamma_B$ (we fix $\Gamma_L$ at $\Gamma_L/D = 0.0157$
and vary $\Gamma_B$ in Fig.~\ref{fig:fig19}), the two scales
$T_K$ and $T_{\rm 1ch}$ are essentially the same, differing by
a factor of three in favor of $T_K$. Upon decreasing $\Gamma_B$
by a factor of four, i.e., $\Gamma_B = \Gamma_L/4$, the order of
scales is reversed, and $T_{\rm 1ch}$ becomes twenty-five-fold
larger than $T_K$. Upon further reducing $\Gamma_B$ to
$\Gamma_B = \Gamma_L/16$, the scale $T_{\rm 1ch}$ exceeds
$T_K$ by four orders of magnitude. Hence the impurity moment
$\vec{\tau}$ is quenched well ahead of the isospin $\vec{S}$
for $\Gamma_L \gg \Gamma_B$, in accordance with the analysis
of sec.~\ref{subsec:large-u}. In this limit one can safely
speak of two successive Kondo effects: first the impurity spin
undergoes one-channel screening by the lead electrons, followed
by two-channel overscreening of the charge fluctuations inside
the box. A similar qualitative picture is recovered for
$\Gamma_B \gg \Gamma_L$, except that the impurity spin is
screened by the box electrons rather than the lead electrons.

Interestingly, the sharp shoulder that characterizes
the capacitance for $\Gamma_L = \Gamma_B$ is lost for
$\Gamma_L \gg \Gamma_B$, and one recovers a capacitance line
shape that closely resembles the $U = 0$ case. In particular,
the onset of the $\ln(T)$ temperature dependence of the capacitance
is pushed back to $T/T_K \sim 0.2$ in Fig.~\ref{fig:fig19}.
The sharp shoulder that develops in $C(0, T)$ for
$\Gamma_L = \Gamma_B \ll U$ is therefore a distinct signature
of the simultaneous quenching of $\vec{S}$ and $\vec{\tau}$
in this case.

\section{Discussion and conclusions}
\label{sec:discussion}

The charging of a quantum box, weakly connected to a lead by
a single-mode point contact, is one of the leading scenarios
for the realization of the two-channel Kondo effect.~\cite{Matveev91}
The main obstacle hampering the observation of this effect in
semiconductor quantum dots stems from the exponential smallness
of the Kondo temperature $T_K$. In order for a fully developed
two-channel Kondo effect to emerge, $k_B T_K$ must significantly
exceed the level spacing inside the box. However, with $T_K$
being exponentially smaller than the charging energy $E_C$,
it is practically impossible to realize a measurable Kondo
scale that exceeds the level spacing in present-day
semiconducting devices.~\cite{ZZW00}

In this paper we proposed a possible remedy to this problem, by
considering a setting in which tunneling between the lead and the
box takes place via a single resonant level. The basic idea
is to exploit the strong energy dependence of the transmission
coefficient through the impurity to achieve nearly perfect
transmission at the Fermi energy, but only small transmission
away from the Fermi energy. In this manner, the large transmission
at the Fermi energy enhances $T_K$, while the small transmission
away from the Fermi energy insures the emergence of a sharp
Coulomb-blockade staircase.~\cite{GM00} This should be contrasted
with the case of an energy-independent transmission coefficient,
where the Coulomb staircase is washed out for perfect
transmission.~\cite{Matveev95}

As seen in Figs.~\ref{fig:fig10} and \ref{fig:fig11}, the
Kondo temperature is indeed dramatically enhanced when the
impurity is tuned close to perfect transmission at the Fermi
energy. Specifically, $T_K$ varies approximately linearly with
the level broadening $\Gamma_L + \Gamma_B$, and is many-fold
larger than the exponentially small Kondo scale obtained
for a tunnel barrier with comparable tunneling matrix elements.
We emphasize that this enhancement of $T_K$ is not restricted
to perfect alignment of the level with the Fermi energy.
Rather, it extends also to a level off resonance with the
Fermi energy, provided that the single-particle transmission
coefficient is large at the Fermi energy. This point is demonstrated
in Fig.~\ref{fig:fig13}, where $k_B T_K/(\Gamma_L + \Gamma_B)$
roughly equals $0.2$, even though $\epsilon_d$ is as large
as $\Gamma_L = \Gamma_B$. Note that the corresponding
transmission coefficient at the Fermi level is indeed large,
being equal to ${\cal T} = 0.8$.

For ${\cal T} \ll 1$, the Kondo temperature is again small.
However, we find a qualitative difference between
$\Gamma_B \ll \Gamma_L$ and $\Gamma_B \gg \Gamma_L$. For
$\Gamma_B \ll \Gamma_L$, the quantum box is only weakly
coupled to the level, and thus to the lead. Accordingly,
the Kondo temperature takes the exponential form
$T_K \propto \exp\!\left [ -\pi^2/2 \sqrt{\cal T} \right]$,
similar to the case of a weak tunnel barrier with a transmission
coefficient equal to ${\cal T}$.~\cite{Matveev91} This result
was derived both for a level at resonance with the Fermi
energy, Eq.~(\ref{T_K:weak_coupling}), and for a level
off resonance with the Fermi energy, see end of
sec.~\ref{sec:Level_off_resonance}.

By contrast, the Kondo scale deviates from the exponential
dependence on $1/\sqrt{\cal T}$, for $\Gamma_B \gg \Gamma_L$.
For $|\epsilon_d| \gg \Gamma_B \gg \Gamma_L$,
for example, $T_K$ depends in a power-law fashion on $\Gamma_L$,
as specified in Eq.~(\ref{T_K_power-law}). For a level at
resonance with the Fermi energy, the ratio $\Gamma_B/\Gamma_L$
defines the crossover from weak ($\Gamma_B/\Gamma_L \ll 1$)
to strong ($\Gamma_B/\Gamma_L \gg 1$) coupling, which
emphasizes the inequivalent roles of $\Gamma_L$ and
$\Gamma_B$ in the two-channel Kondo effect that develops.

Any practical realization of a resonant level will necessarily
involve an on-site Coulomb repulsion $U$ on the level. Such
an interaction couples the two spin orientations, questioning
the development of the two-channel Kondo effect. In this paper
we focused on a symmetric level, $U + 2\epsilon_d = 0$, which is
easier to study numerically since both the degeneracy point and
the mid-charge point are pinned at $V_B = 0$. For an asymmetric
level, $U + 2\epsilon_d \neq 0$, the degeneracy point (assuming it
exists) is shifted away from $V_B = 0$, and is no longer identified
with the mid-charge point. For $U = 0$, this was explicitly shown
to be the case in Figs.~\ref{fig:fig12} and \ref{fig:fig13}.

As seen in Fig.~\ref{fig:fig15}, the two-channel Kondo effect
is robust against the inclusion of an on-site repulsion, for
a symmetric level. For $\Gamma_L = \Gamma_B$, the Kondo scale
extracted from the slope of the $\ln(T)$ diverging term in
the capacitance is actually enhanced by a moderately large
repulsion $U$ (see Fig.~\ref{fig:fig16}), although the onset
of the logarithmic temperature dependence is pushed down to
lower temperature. For $\Gamma_L < \Gamma_B$ and intermediate
$U$, there are clearly two distinct energy scales governing
the low-temperature capacitance: a thermodynamic crossover scale
$k_B T_0$, below which the $\ln(T)$ temperature dependence
sets in, and a low-temperature scale $k_B T_K$, which fixes
the slope of the $\ln(T)$ diverging term. This differs
from the conventional two-channel Kondo Hamiltonian, where
$T_0 \sim 0.2 T_K$ and $T_K$ are essentially the same.

For a large on-site repulsion, $U \gg \Gamma_L + \Gamma_B$,
the Kondo temperature is again exponentially small, this time
in $U /(\Gamma_L + \Gamma_B)$, see Figs.~\ref{fig:fig16}
and \ref{fig:fig17}. For $\Gamma_L \gg \Gamma_B$, one can
qualitatively understand this result along the following
lines. As seen in Fig.~\ref{fig:fig19}, for
$\Gamma_L \gg \Gamma_B$ there is a sequence of Kondo effects:
first the local moment on the level undergoes single-channel
screening by the lead electrons, followed by two-channel
overscreening of the charge fluctuations inside the box.
Below the single-channel Kondo temperature
$T_{\rm 1ch} \propto \exp \left [ -\pi U/8 \Gamma_L \right]$,
the spectral function of the $d$ electrons consists of two
high-energy resonances at $\pm U/2$, and a narrow Abrikosov-Suhl
resonance of width $T_{\rm 1ch}$ and height $1/\pi \Gamma_L$, which
is pinned at the chemical potential.~\cite{Hewson93} It is the
latter peak that serves as the effective lead density of states,
$\rho^{\rm eff}_{L}(\epsilon)$ of Eq.~(\ref{Lorentzian}),
available for screening the charge fluctuations inside the box.
Substituting $T_{\rm 1ch}$
in for $D_L$ in Eq.~(\ref{D^ast}), the pre-exponential
factor in Eq.~(\ref{T_K:weak_coupling}) is reduced by
a factor of $\sqrt{T_{\rm 1ch}/{\Gamma_L}}$, yielding a
Kondo temperature which is exponentially small in both
$U/\Gamma_L$ and $\sqrt{\Gamma_L/\Gamma_B}$. Obviously,
the above picture overlooks the possible relevance of
higher order interaction terms generated upon the
screening of the $d$ local moment.

In semiconducting devices, one can realize a tunable level
using an ultrasmall quantum dot, whose charging energy $U$
is bound to exceed that of the quantum box. Since the
bandwidth $D$ in the effective Hamiltonian of
Eqs.~(\ref{H_L})--(\ref{H_tun}) is of the order of $E_C$, this
dictates the hierarchy $U > D \gg \Gamma_L, \Gamma_B$.
Hence, not much can be gained from a symmetric level, as
the associated Kondo temperature $T_K$ is exponentially small
in $U/(\Gamma_L + \Gamma_B) \gg 1$. Instead, one would like
to tune the level to the mixed-valent regime, where $T_K$ is
expected to be of the order of the level broadening.

As emphasized above, treatment of such an asymmetric level is
complicated by the need to accurately locate the position of
the degeneracy point, which is no longer pinned at $V_B = 0$,
and does not coincide with the mid-charge point. In fact, one
cannot entirely rule out the possibility that the two-channel
Kondo effect is unstable against particle-hole asymmetry for
an interacting level, as recently implied by Le Hur and
Simon.~\cite{LHS02} Using perturbative RG and an analogy
to a related model of two capacitively coupled quantum
dots,~\cite{BZHHvD03} these authors argued that the two-channel
Kondo is lost in the local-moment regime, when a stable local
moment is formed on the level.~\cite{comment_on_LHS02}
Although no explicit distinction was made between a
particle-hole symmetric and an asymmetric level, the
analysis of Le Hur and Simon implicitly assumed an
asymmetric level, by taking the coupling $V$ to be nonzero
(see Ref.~\onlinecite{LHS02}). For $V = 0$, there are
qualitative changes to the RG flow of Le Hur and Simon.
Indeed, Figs.~\ref{fig:fig15}--\ref{fig:fig19} unambiguously
establish the emergence of the two-channel Kondo effect for
$2\epsilon_d + U = 0$, including in the local-moment
regime $U \gg \Gamma_L, \Gamma_B$.

For weak coupling, $\Gamma_L \gg \Gamma_B, |\epsilon_d|, U$,
one can analytically see that particle-hole asymmetry does not
play any role in the emergence of the two-channel Kondo effect.
To this end, we extend the analysis of sec.~\ref{subsec:weak-coupling}
to an asymmetric level, $2\epsilon_d + U \neq 0$. The effect of
a nonzero $\Delta \epsilon = \epsilon_d + U/2$ is to supplement
the Hamiltonian terms of Eqs.~(\ref{H_energy}) and (\ref{H_U_energy})
with an additional potential-scattering term for the
$a^{\dagger}_{\epsilon L \sigma}$ degrees of freedom:
\begin{equation}
{\cal H}_{PS} = \Delta \epsilon
                 \sum_{\sigma}
                 \int \!\! d\epsilon_1\!\!
                 \int \!\! d\epsilon_2 
		    v(\epsilon_1, \epsilon_2)
                    :\!a^{\dagger}_{\epsilon_1 L \sigma}
                       a_{\epsilon_2 L \sigma}\!\!:
\label{H_PS}
\end{equation}
with
\begin{equation}
v(\epsilon_1, \epsilon_2) =
       \sqrt{\rho_L^{\rm eff}(\epsilon_1) \rho_L^{\rm eff}(\epsilon_2)} .
\end{equation}
Replacing $\rho^{\rm eff}_L(\epsilon)$ with the symmetric
rectangular DOS of Eq.~(\ref{rho_box}), the
potential-scattering term of Eq.~(\ref{H_PS}) reads
\begin{equation}
{\cal H}_{PS} = \frac{\Delta \epsilon}{\pi \Gamma_L}
                 \sum_{\sigma}
                 \int_{-D_L}^{D_L} \!\! d\epsilon_1\!\!
                 \int_{-D_L}^{D_L} \!\! d\epsilon_2\!
                    :\!a^{\dagger}_{\epsilon_1 L \sigma}
                       a_{\epsilon_2 L \sigma}\!\!: \ ,
\end{equation}
which is just the sum of the two Hamiltonian terms of
Eq.~(\ref{V_pm_terms}) with $D_m \to D_L$ and
$\tilde{V}_{+} = \tilde{V}_{-} = \Delta \epsilon/(2 \pi \Gamma_L)$.
Aside from renormalizing the voltage $V_B$, the addition of such
a Hamiltonian term does not affect the low-energy physics.
The system continues to show the two-channel Kondo effect,
albeit at a shifted position of the degeneracy point. This
differs from the conclusion of Le Hur and Simon for the
local-moment regime.

Away from weak coupling, a full-scale numerical effort is
required to resolve the effect of particle-hole asymmetry
on the emergence of the two-channel Kondo effect. Such a
study is currently under way. Our preliminary results for
the mixed-valent regime (setting $\epsilon_d$ equal to zero)
support the conclusion that the two-channel Kondo effect
is robust against particle-hole asymmetry for an interacting
level. Moreover, there are indications that $T_K$ remains
significantly enhanced in the mixed-valent regime also for
$U$ several times larger than $D \sim E_C$, as is the case
in realistic semiconducting devices. A detailed analysis
of this physically relevant regime will be presented in a
forthcoming publication.\vspace{10pt}

\section*{ACKNOWLEDGMENTS}\vspace{-5pt}

We are grateful to Daniel Cox for many illuminating discussions,
and to the members of the condensed matter theory group at UC Davis
for their warm hospitality during the early stages of this work.
E.L. and A.S. were supported in part by the Centers of Excellence
Program of the Israel science foundation, founded by The Israel
Academy of Science and Humanities, and by the Niedersachsen-Israel
foundation. F.B.A. was supported in part by DFG grant AN 275/2-1.


\begin{thebibliography}{99}
\bibitem{comment_on_rho_L^eff} We use the convention by which
       $\rho_L^{eff}(\epsilon)$ is normalized to unity.
\bibitem{comment_on_D} Recall that $D$ itself is of the order of the
       charging energy $E_C$, which serves as the high-energy cutoff
       for the mapping of Eqs.~(\ref{H_L})--(\ref{H_tun}).
\bibitem{comment_on_dos} One can consider more realistic density of
       states for the lead and the quantum box. However, apart from
       a possible shift in the position of the degeneracy point, we
       do not expect any significant modifications to the physical
       picture.
\bibitem{comment_on_numerics} We computed the capacitance
       $C(V_B, T)$ by numerically differentiating the calculated
       excess charge, $\delta Q(V_B, T)$. This was done in the
       following manner. At the degeneracy point, $V_B = 0$,
       we used $C(0, T) = \delta Q(v, T)/v$, where $v$ was
       taken to be a small voltage of the order of
       $e v/D \sim 10^{-9} - 10^{-11}$. For temperatures above
       $T_{sp} = (e v)^2/ k_B^2 T_K$, this procedure proved
       very stable, with only minute variations as $ev/D$ was
       lowered from $10^{-9}$ down to $10^{-13}$. Away from
       the degeneracy point we used $C(V_B, T) =
       [\delta Q(V_B + v, T) - \delta Q(V_B - v, T)]/2v$,
       with $v = 0.1 V_B$.
\bibitem{comment_on_A_Lambda} When comparing the NRG to expressions
       derived in the continuum limit, one generally needs to rescale
       the hybridization widths in the continuum limit according
       to $\Gamma_L \to \Gamma_L/A_{\Lambda}$ and
       $\Gamma_B \to \Gamma_B/A_{\Lambda}$, with
       $A_{\Lambda} = [(\Lambda + 1)/2(\Lambda -1)] \ln\Lambda$
       (see Ref.~\onlinecite{KWW80a}). This correction, which
       accounts for the energy discretization used in the NRG, does
       not enter the exponent in Eq.~(\ref{T_K:weak_coupling}), as
       the latter depends on the ratio $\Gamma_L/\Gamma_B$. The NRG
       discretization factor only modifies the pre-exponential
       coefficient in Eq.~(\ref{T_K:weak_coupling}).
\bibitem{comment_on_p-h} Explicitly, the parameter regime
       $\epsilon_d < 0$ is obtained from $\epsilon_d > 0$
       by the particle-hole transformation
       $c \to c^{\dagger}, d \to d^{\dagger}$, and
       $(S_x, S_y, S_z) \to (-S_x, S_y, -S_z)$, which maps
       $\epsilon_d$, $V_B$, and $\delta Q$ onto $-\epsilon_d$,
       $-V_B$, and $-\delta Q$, respectively.
\bibitem{comment_on_LHS02} The treatment of Le Hur and Simon closely
       resembles that of sec.~\ref{subsec:large-u}, except that the
       condition $2\epsilon_d + U = 0$ is relaxed. Setting
       $2\epsilon_d + U = 0$ in the expressions of
       Ref.~\onlinecite{LHS02} reveals several differences
       in the effective low-energy Hamiltonian produced by
       the Schrieffer-Wolff transformation as compared to
       Eqs.~(\ref{H_SW})--(\ref{V_0_bare}). These differences
       stem from the fact that we apply the Schrieffer-Wolff
       transformation to the Hamiltonian of
       Eqs.~(\ref{H_L})--(\ref{H_tun}) and (\ref{Hubbard}),
       in which all excited charge configurations in the box
       have been projected out. By contrast, Le Hur and Simon
       retain the effect of the excited charge configurations,
       as is required in the limit where
       $E_C \ll -\epsilon_d, U + \epsilon_d$.
\end{thebibliography}
\end{document}